\PassOptionsToPackage{dvipsnames}{xcolor}
\PassOptionsToPackage{hyphens}{url}
\documentclass[sigplan,screen]{acmart}

\startPage{1}

\setcopyright{rightsretained}

\copyrightyear{2023}
\acmYear{2023}
\setcopyright{rightsretained}
\acmConference[PPoPP '23]{The 28th ACM SIGPLAN Annual Symposium on Principles and Practice of Parallel Programming}{February 25-March 1, 2023}{Montreal, QC, Canada}
\acmBooktitle{The 28th ACM SIGPLAN Annual Symposium on Principles
and Practice of Parallel Programming (PPoPP '23), February 25-March 1,
2023, Montreal, QC, Canada}
\acmDOI{10.1145/3572848.3577478}
\acmISBN{979-8-4007-0015-6/23/02}

\usepackage{amsmath,amsfonts}
\usepackage{algorithmic}
\usepackage{graphicx}
\usepackage{textcomp}
\usepackage[ruled,vlined,linesnumbered]{algorithm2e}
\usepackage{subfiles}
\usepackage{bm}
\usepackage{soul}
\usepackage[utf8]{inputenc}
\usepackage[T1]{fontenc}
\usepackage{url}
\usepackage{hyperref} 
\usepackage{booktabs}
\usepackage{amsfonts}
\usepackage{nicefrac}
\usepackage{microtype}      
\usepackage{listings}
\usepackage{enumitem}
\usepackage[noabbrev]{cleveref}
\usepackage{stfloats}
\usepackage{varwidth}
\usepackage{multicol}
\usepackage{multirow}
\usepackage{makecell}
\usepackage{float} 
\usepackage{listings}
\usepackage{caption}
\usepackage{subcaption}
\lstdefinestyle{cuda}{
	belowcaptionskip=1\baselineskip,
	breaklines=true,
	xleftmargin=\parindent,
	language=C,  
	showstringspaces=false,
	basicstyle=\footnotesize\ttfamily,
	keywordstyle=\bfseries\color{PineGreen},
	commentstyle=\color{white!40!black},
	identifierstyle=\color{NavyBlue},
	stringstyle=\color{orange},
	escapeinside={(*@}{@*)},
	morekeywords={shared, is, to},
	numbers=left,
	morekeywords={[2]{copy,syncthreads,nnz}},
	keywordstyle={[2]{\color{Purple}}},
}     
\def\BibTeX{{\rm B\kern-.05em{\sc i\kern-.025em b}\kern-.08em
    T\kern-.1667em\lower.7ex\hbox{E}\kern-.125emX}}
    
\newcommand{\bmmc}[1]{\bm{\mathcal{#1}}}
\newcommand{\RR}[1]{\mathbb{R}^{#1}}

\DeclareMathOperator*{\argmin}{arg\,min}

\hypersetup{pdfauthor={L. Xiang}, pdftitle={TDC}}

\begin{document}
\title[\textsf{TDC}]{TDC: Towards Extremely Efficient CNNs on GPUs via Hardware-Aware Tucker Decomposition}

\newcommand{\AFFIL}[3]{%
    \affiliation{%
        \institution{\small #1}
        \city{\small #2}\state{\small #3}
    }
    }

\newcommand{\iu}{Indiana University}
\newcommand{\wsu}{Washington State University}
\newcommand{\metautah}{Meta and University of Utah}
\newcommand{\ru}{Rutgers University}
\newcommand{\uu}{University of Utah}

\newcommand{\WSU}{\AFFIL{\wsu}{Pullman}{WA}}
\newcommand{\IU}{\AFFIL{\iu}{Bloomington}{IN}}
\newcommand{\MU}{\AFFIL{\metautah}{}{}}
\newcommand{\RU}{\AFFIL{\ru}{New Brunswick}{NJ}}
\newcommand{\UU}{\AFFIL{\uu}{Salt Lake City}{UT}}

\author{Lizhi Xiang}{\UU}
\email{xianglizhi456@gmail.com}
\authornote{Both authors contributed equally to the paper.}
\authornote{This work was done when Lizhi Xiang, Chengming Zhang, and Dingwen Tao were with Washington State University.}

\author{Miao Yin}{\RU}
\email{miao.yin@rutgers.edu}
\authornotemark[1]

\author{Chengming Zhang}{\IU}
\email{czh5@iu.edu}
\authornotemark[2]

\author{Aravind Sukumaran-Rajam}{\MU}
\email{aravindsr@meta.com}

\author{P. Sadayappan}{\UU}
\email{saday@cs.utah.edu}

\author{Bo Yuan}{\RU}
\email{bo.yuan@soe.rutgers.edu}

\author{Dingwen Tao}{\IU}
\email{ditao@iu.edu}
\authornote{Dingwen Tao is the corresponding author.}
\authornotemark[2]

\renewcommand{\shortauthors}{Xiang et~al.}

\begin{abstract}
Tucker decomposition is one of the SOTA CNN model compression techniques. However, unlike the FLOPs reduction, we observe very limited inference time reduction with Tucker-compressed models using existing GPU software such as cuDNN. To this end, we propose an efficient end-to-end framework that can generate highly accurate and compact CNN models via Tucker decomposition and optimized inference code on GPUs. Specifically, we propose an ADMM-based training algorithm that can achieve highly accurate Tucker-format models. We also develop a high-performance kernel for Tucker-format convolutions and analytical performance models to guide the selection of execution parameters. We further propose a co-design framework to determine the proper Tucker ranks driven by practical inference time (rather than FLOPs). Our evaluation on five modern CNNs with A100 demonstrates that our compressed models with our optimized code achieve up to 2.21$\times$ speedup over cuDNN, 1.12$\times$ speedup over TVM, and 3.27$\times$ over the original models using cuDNN with \textcolor{black}{at most} 0.05\% accuracy loss.
\end{abstract}

\keywords{Convolutional neural network, inference, GPU, performance, model compression.}  

\ccsdesc[500]{Computing methodologies~Massively parallel algorithms}
\ccsdesc[300]{Software and its engineering~Source code generation}
\ccsdesc[500]{Computing methodologies~Neural networks}

\maketitle

\setlength{\textfloatsep}{6pt}
\setlength\abovecaptionskip{3pt}
\setlength{\abovedisplayskip}{2pt}
\setlength{\belowdisplayskip}{2pt}
\setlength{\abovedisplayshortskip}{2pt}
\setlength{\belowdisplayshortskip}{2pt}

\section{Introduction}
\label{sec:intro}
Not only are deep neural networks such as convolutional neural networks (CNNs) critical to traditional AI tasks \cite{he2020lightgcn, collins2001convolution}, but their applications in science domains are also growing \cite{cherukara2020ai, kurth2018exascale, mathuriya2018cosmoflow}.
However, in recent years, demands to improve the inference performance of CNNs have never been satisfied. Prior works approach faster and more efficient CNNs from different aspects, such as weight pruning \cite{ma2020pconv}, quantization \cite{feng2021apnn,jin2019deepsz}, and kernel factorization \cite{povey2018semi,yin2021towards}. 

Pruning techniques are proposed to remove the weight redundancy and produce an irregular sparse model to reduce the computational cost and memory bandwidth requirements of inference. However, this requires specialized sparsity-aware accelerators with significant effort in new hardware design and fabrication. 
Moreover, some researchers proposed to create a regular sparsity pattern through hardware-aware pruning algorithms \cite{ma2020pconv, dong2020rtmobile, zhang2021clicktrain}, but their compression ratio is largely limited by the enforced sparsity patterns.

Quantization is to reduce the number of bits to store the model weights but faces two main challenges: 
\underline{(1)} The quantized CNNs usually require arbitrary precisions (e.g., 4-bit weight), while widely used accelerators such as GPUs only support a limited range of precisions (e.g., half precision).
\underline{(2)} The compression ratio provided by quantization is still insufficient, considering many large CNN models.

Due to the above limitations, recent CNN model compression studies have mainly focused on a new direction---building the compact models using kernel factorization such as matrix/tensor decomposition. Specifically, matrix/tensor decomposition can represent a large matrix/tensor with the combination of multiple small matrices/tensors. Correspondingly, the number of the required representation parameters can be significantly reduced. 
However, matrix decomposition needs to convert high-order tensors to matrices and then perform matrix decomposition, causing loss of spatial information, whereas high-order tensor decomposition achieves much higher accuracy \cite{yin2021towards}.
Thus, various compact CNN models have been developed using different tensor decomposition approaches \cite{garipov2016ultimate,wang2018wide,kim2015compression}. Among those efforts, Tucker decomposition (TKD) \cite{tucker1966some} finds its strengths in minimal modification of the hardware support, ultra-high compression ratio, and lower accuracy loss for CNN acceleration. 

We identify three major challenges in accelerating TKD-compressed CNNs on state-of-the-art GPUs. \par
\begin{figure*}[t]
    \centering
    \includegraphics[width=0.75\linewidth]{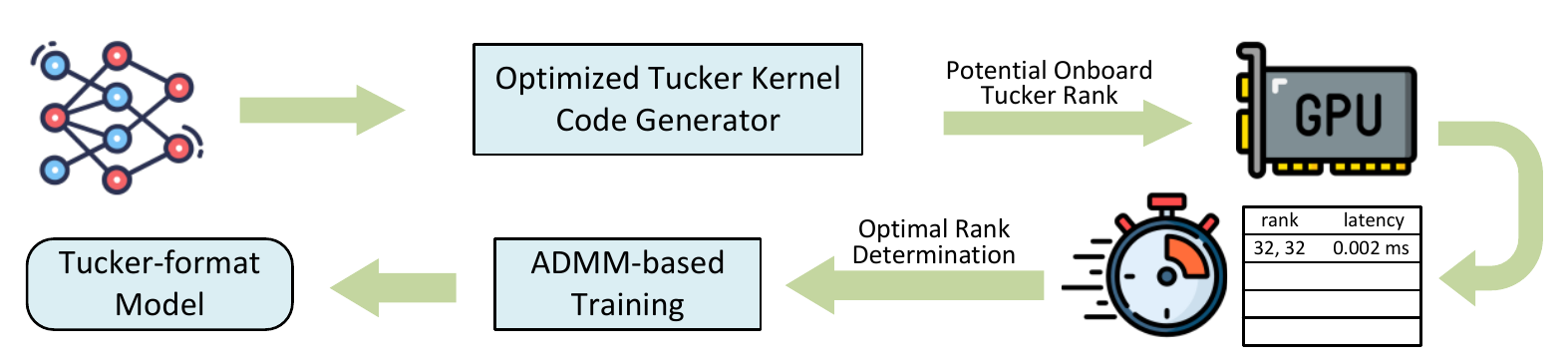}
    \vspace{-2mm}
    \caption{Overview of our \textsf{TDC} framework for generating TKD-compressed CNN models with high-performance inference code on GPUs.}
    \vspace{-4mm}
    \label{fig:overall}
\end{figure*}
\textbf{Lack of hardware-aware TKD algorithms for CNN acceleration.}
Once the target hardware platform is selected, Tucker ranks are the most important parameters impacting the practical latency. Unfortunately, existing TKD compression methods \cite{kim2015compression,gusak2019automated} select ranks via either laborious efforts or iterative search, none of which are hardware aware.

\textbf{Lack of software-aware TKD convolution algorithms for CNN acceleration.}
Tucker-based convolution decomposition can greatly decrease the computation FLOPs by reducing the number of the input channels and output channels of the original convolution layer. However, it is hard to translate the FLOPS reduction to real performance increment especially for small batch size such as one\footnote{A batch size of 1 is important for ML applications that require real-time responses \cite{batch-size-one}, such as navigation systems in autonomous driving. As another example, the ability to auto-fill in suggestions when a retail customer enters a product name into an e-commerce site is required for optimal response.
}. The root cause is that Tucker-based convolutions suffer from severe resource under-utilization when executed using existing software like cuDNN. 
For example, our evaluation (will be shown in Figure \ref{fig:A100-end2end}) illustrates that TKD-compressed ResNet18 model (with 2.7$\times$ FLOPs reduction) using cuDNN only achieves 1.47$\times$ speedup over the original model 
on A100 GPU. 

\textbf{Lack of performance model for tuning tucker-format convolutions on GPUs.}
Popular software for high performance convolutions such as TVM can generate fast convolution code by tuning hyper-parameters via machine learning (ML) approaches on a given set of convolution templates. However, those computation schemes do not work perfectly on the tucker-format convolutions on GPUs (will be detailed in Section \ref{sec:tvm-limit}). To solve this issue, an analytical performance model to guide the parameter tuning for tucker-format convolutions on GPUs is urgently needed.\par
\textbf{Lack of performance-driven frameworks for highly efficient and accurate CNN inference on GPUs.}
Existing model compression methods are primarily driven by FLOPs reductions rather than real runtime reductions. However, it is still unclear how to design an end-to-end compression framework (including training highly accurate TKD-compressed models and building optimized inference code) to reduce the model size to the sweet spot where further compression cannot provide any performance benefit. 

To this end, we propose an end-to-end framework, called \underline{T}ucker \underline{D}ecomposition \underline{C}onvolution (\textsf{TDC}), that can generate highly accurate and compact TKD-compressed CNN models with optimized C++/CUDA code, which can be easily deployed on GPUs for high-performance inference, as illustrated in Figure \ref{fig:overall}. 
First, we propose an Alternating Direction Method of Multipliers (ADMM) based training algorithm for TKD-format convolutions and incorporate inference performance into the model generation. 
Second, we propose a new computation scheme for Tucker-format convolutions on GPUs. We also build a performance model to guide the selection of tiling sizes for our convolution kernel rather than performing an exhaustive search of the best tiling sizes.
Third, we propose a co-design approach that can automatically determine the proper Tucker ranks based on a target speedup of inference performance. Based on the determined ranks, we can train a highly accurate TKD-compressed model and generate the optimized code for inference on GPUs. 
The main contributions of this paper are summarized as follows:
\begin{itemize}[topsep=0pt,partopsep=0pt,parsep=0pt]
    \item We propose an algorithm-hardware co-designed, TKD-based compression framework that can achieve inference acceleration on GPUs with high model accuracy.
    \item We redesign the ADMM-based training algorithm for TKD-compressed models.
    \item We develop a new kernel for Tucker-format convolutions and a performance model for tiling size selection. 
    \item We develop an approach to choose the proper Tucker ranks based on practical inference runtime and plug them into the training phase for model generation. 
    \item 
    Evaluation results on five modern CNN models with ImageNet demonstrate that our solution can speed up the end-to-end inference on A100 by 1.26$\sim$2.21$\times$ over cuDNN and 1.03$\sim$1.12$\times$ over TVM on the tested models. Moreover, compared with the original models, our compressed models achieve up to 3.27$\times$ speedup with at most 0.05\% accuracy loss (even higher accuracy). 
\end{itemize}

In Section \ref{sec:background}, we introduce CUDA architecture, TKD-based model compression, and GPU convolution.  
In Section \ref{sec:tucker-algorithm}, we present our hardware-aware tucker decomposition for CNN acceleration. In Section \ref{sec:micor-kernel}, we discuss the details of our convolution scheme. In Section \ref{sec:co-design}, we present our new co-design framework.  In Section \ref{sec:evaluation}, we show our evaluation results.  In Section \ref{sec:conclusion}, we conclude this work and discuss future work. 

\section{Background}
\label{sec:background}
In this section we present the background about CUDA GPU architecture, tensor decomposition for model compression, and high-performance software for GPU convolutions. 

\subsection{CUDA Architecture}
The thread is the basic programmable unit that allows GPU programmers to use  massive numbers of CUDA cores. CUDA threads are grouped at different levels such as warp, block, and grid.
Specifically, a group of 32 threads is called a \textit{warp}. All threads in the same warp will execute the same instruction. However, if different threads in a warp follow different control paths, some threads are masked from performing any useful work. This situation is called a \textit{warp divergence}, which is one of the fundamental factors that limit the performance of GPUs. Multiple warps are combined to form a thread \textit{block}, and the set of thread blocks is called the \textit{grid}. 



\subsection{Related Work on Tensor Decomposed DNN}
\textbf{Tensor Decomposition for DNNs.} Higher-order tensor decomposition is an advanced technique to explore multi-dimensional linear correlations on the weight kernels via directly decomposing high-order kernels to multiple small 
tensor cores. 
Multiple tensor decomposition formats have been successfully applied in CNN compression. In prior work \cite{lebedev2014speeding, phan2020stable}, CP decomposition is adopted to factorize weight tensors to reduce computational and storage costs of convolutional layers. Tensor-train (TT) \cite{oseledets2011tensor} and its variant tensor-ring (TR) \cite{zhao2016tensor} decomposition are also commonly used low-rank compression methods for CNNs  \cite{novikov2015tensorizing,garipov2016ultimate,yinadmm2021}. In addition, \cite{kim2015compression,gusak2019automated} propose to compress and accelerate CNN models via using Tucker decomposition, and \cite{liang2021automatic} further develops a mixed CP and Tucker method to improve the model accuracy. 
However, 
these existing works still face several challenges for efficient CNN acceleration. 

\textbf{Limitations of CP-based Method \cite{phan2020stable}.} 
In general, CP decomposition suffers two limitations for high performance CNN acceleration. At the algorithm level, the CP method, by its nature, has an inherent instability problem, thereby causing inferior model accuracy. Though the state-of-the-art (SOTA) CP method aims to mitigate this challenge, its accuracy is still unsatisfying. At the hardware level, the ranks of different modes in CP are always identical when decomposing a kernel tensor, which hinders the adjustment of the read-write loads in memories by changing the ranks ratio. 

\textbf{Limitations of TT-based Method \cite{yinadmm2021}.}
The state-of-the-art TT compression \cite{yinadmm2021} has two inherent drawbacks. First, as stated in \cite{garipov2016ultimate}, to compress a 4-D weight tensor, TT needs to first convert the original convolution to a huge matrix-matrix multiplication, and then uses TT to transform the multiplication to a TT-FC layer \cite{novikov2015tensorizing}. In such a conversion, important spatial information in filter dimensions (RxS) is lost and the convolution process no longer exists, thereby causing non-negligible accuracy drop. Second, according to tensor theory, when decomposing a 4-D weight tensor, with the similar rank setting TT brings lower computational cost reduction than other methods such as Tucker.


\textbf{Limitations of Tucker-based Method \cite{gusak2019automated}.} The state-of-the-art Tucker work \cite{gusak2019automated} still has unsatisfactory accuracy performance when compressing modern CNNs (e.g., ResNet) on ImageNet dataset (will be shown in Table \ref{tab:accuracy}). Moreover, the rank selection of \cite{gusak2019automated} is not performed in a hardware-friendly way, thus limiting its runtime performance.


\textbf{Tucker Method for RNN Compression.} Tucker decomposition can also be used for compressing matrix-vector-multiplication-centered models, e.g., recurrent neural networks (RNNs). To that end, we can directly reshape the weight matrix to a high-order tensor, then decompose it into Tucker format. 
Under the Tucker format, the original matrix-vector multiplication is converted to
a series of matrix multiplications in the execution. 
Considering that matrix multiplication is well optimized in existing software stacks and the Tucker-based CNN kernel optimization is still under-investigated, we focus on CNNs in this work.

\subsection{High-Performance Convolutions on GPUs}

Software for high-performance convolutions can be categorized into two classes: \underline{(1)} library-based implementations and \underline{(2)} code generator-based implementations. The most famous convolution library is cuDNN \cite{chetlur2014cudnn}, which is widely used in open-source frameworks such as TensorFlow \cite{abadi2016tensorflow} and PyTorch \cite{paszke2019pytorch}. Specifically, a library like cuDNN can take arbitrary convolution problem size as input and is very easy to use. However, the performance of library-based implementation is often less than the code generator-based approaches. This is because library-based implementations are optimized to handle a range of problem sizes, whereas the code generator can be targeted to optimize a given problem size, thus potentially achieving higher performance. Since the problem size is known, the code generator can precisely control factors that affect performance, such as tile sizes, thread coarsening levels, and memory placement strategies. For example, TVM \cite{chen2018tvm} is a well-known code generation framework that generates fast convolution code by tuning hyper-parameters on given convolution templates. However, unlike TVM, which relies on heavy auto-tuning and ML processes, our code-generation-based approach employs analytical performance models to efficiently select the best execution parameters for both high performance and accuracy. 

\section{Basics of TKD-based Convolution}
\label{sec:basic}
In this paper, bold calligraphic letters, bold capital letters and bold lowercase letters represent high-order tensors, matrices and vectors, respectively, e.g., $\bmmc{A}, \bm{A}, \bm{a}$; non-bold letter with indices $\mathcal{A}(i_1,\cdots,i_k)$ denotes the $(i_1,\cdots,i_k)$-th entry of a $k$-th order tensor $\bmmc{A}$; $A(i_1, i_2)$ and $a(i)$ denote the corresponding entry of the matrix $\bm{A}$ and vector $\bm{a}$, respectively. All the notations are summarized in Table \ref{table:notations}.
\begin{table}[h]
\vspace{-4mm}
    \caption{Notation summary.}
\centering
\footnotesize
\begin{tabular}{|l|l|l|l|}
    \hline
    $C$             & \# i/p channels      & $N$       & \# o/p channels       \\ \hline
    $H$             & image height & $W$             & image width \\ \hline
    $R$             & stencil height  & $S$             & stencil width   \\ \hline
    $\bmmc{X}$ & input tensor    & $\bmmc{Y}$ & output tensor   \\ \hline
    $\bmmc{K}$ & kernel tensor   &   &\\ \hline
    \end{tabular}

    \label{table:notations}
\vspace{-4mm}
\end{table}

Give a $k$-th order tensor $\bmmc{A}\in\RR{N_1\times\cdots\times N_k}$, with Tucker decomposition, it can be generally represented as
\begin{equation*}
\small
\label{eq:tucker}
\mathcal{A}(i_1,\cdots,i_k)=\sum_{d_1,\cdots,d_k}^{D_1,\cdots,D_d} \mathcal{C}(d_1,\cdots,d_k) \cdot U_1(i_1,d_1)\cdots U_k(i_k, d_k).
\end{equation*}
$\bmmc{C}\in\RR{D_1\times\cdots\times D_d}$ is called \textit{core tensor}, $\{\bm{U}_i\in\RR{N_i\times D_i}\}_{i=1}^k$ are called \textit{factor matrices}, $[D_1,\cdots,D_d]$ are called \textit{Tucker ranks}.

\begin{figure}[t]
    \centering
    \includegraphics[width=0.85\linewidth]{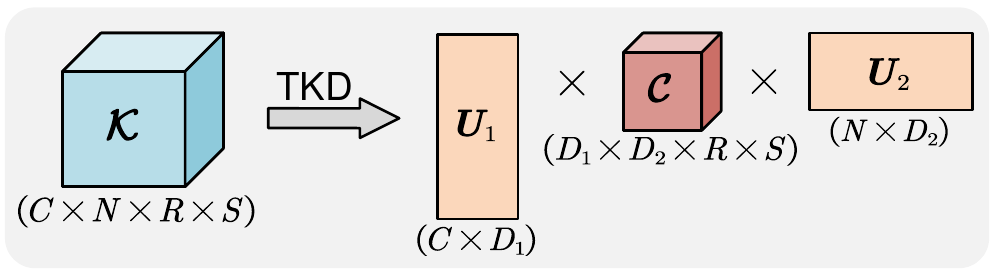}
    \caption{The original 4D convolution kernel is converted to three components in a Tucker-format convolution layer.}
    \label{fig:tk-decomp}
\end{figure}

For compressing a convolutonal layer with kernel tensor $\bmmc{K}\in\RR{C\times N\times R\times S}$ using Tucker decomposition, to preserve the spacial information, we only decompose the first mode and the second mode, i.e., the output and the input channel. 
Hence, kernel $\bmmc{K}$ can be decomposed as
\begin{equation}\label{eq:tkd2}
\mathcal{K}(c,n,r,s)=\sum_{d_1,d_2}^{D_1,D_2}\mathcal{C}(d_1,d_2,r,s)U_1(c,d_1)U_2(n,d_2),
\end{equation}
where $\bmmc{C}\in\RR{C\times N\times D_1\times D_2}, \bm{U}_1\in\RR{C\times D_1}, \bm{U}_2\in\RR{N\times D_2}$ are the compressed components to store and compute in the Tucker-format convolution layer. The decomposed components are illustrated in Figure \ref{fig:tk-decomp}.

With the above Tucker-format decomposed kernel, a convolution layer can be performed with three small consecutive convolutions. Specifically, for an input tensor $\bmmc{X}$ $\in$ $\RR{H\times W\times C}$, the output tensor $\bmmc{Y}$ $\in$ $\RR{H'\times W'\times N}$ is computed by
\vspace{-1.5mm}
\begin{small}
\begin{align}
\mathcal{Z}_1(h,w,d_1)&=\sum_{c=1}^{C}\mathcal{X}(h,w,c)U_1(c,d_1),\label{eq:first_conv}\\
\mathcal{Z}_2(h',w',d_2)&=\sum_{r=1}^R\sum_{s=1}^S\sum_{d_1}^{D_1}\mathcal{Z}_1(h,w,d_1)\mathcal{C}(d_1,d_2,r,s),\label{eq:core_conv}
\end{align}
\end{small}
\begin{small}
\begin{align}
\mathcal{Y}(h',w',n)=&\sum_{d_2}^{D_2}\mathcal{Z}_2(h',w',d_2)U_2(d_2,n),\label{eq:last_conv}
\end{align}
\end{small}
where $h=h'+r-1, w=w'+s-1$. 

\begin{figure}[t]
    \centering
    \includegraphics[width=.95\linewidth]{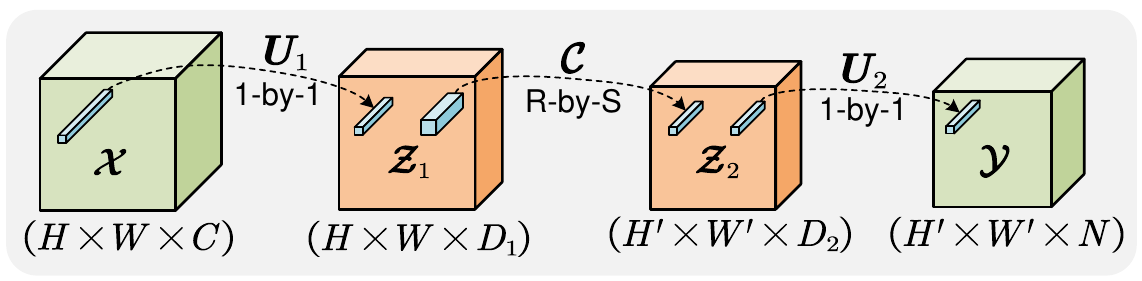}
    \caption{Illustration of Tucker-format convolution.}
    \label{fig:tk-conv}
    \vspace{-2mm}
\end{figure}

The above Tucker-format convolution is mathematically equivalent to the original convolution.
Figure \ref{fig:tk-conv} illustrates the Tucker-format convolution. 
The first operation Eq. (\ref{eq:first_conv}) is essentially a channel-wise 1$\times$1 convolution that transforms original channels $C$ to the smaller latent channels $D_1$. Then the computation Eq. (\ref{eq:core_conv}) can be considered as the regular 2D convolution with the same kernel size $R,S$ from channels $D_1$ to $D_2$ (called \textit{core convolution}). The last computation Eq. (\ref{eq:last_conv}) is also a channel-wise 1$\times$1 convolution that transforms the latent channels $D_2$ to the original output channels $N$.

\textbf{Benefits from hardware perspective.} Note that regular $R$-by-$S$ convolution is generally more sensitive to size than one-by-one channel-wise convolution. With this Tucker-format convolution, one can customize the second convolution by defining $D_1$ and $D_2$ as the hardware such as GPU friendly sizes, then use channel-wise convolution to complement the unbalanced custom channel sizes.

\textbf{Computation and storage complexity.} The overall storage complexity is number of parameters of factor matrices $\bm{U}_1,\bm{U}_2$ and core tensor $\bmmc{C}$, i.e., $C \cdot D_1+N \cdot D_2+R \cdot S \cdot D_1 \cdot D_2$. The overall computation complexity is also the sum of FLOPs of the three convolutions, i.e., $H \cdot W \cdot C \cdot D_1+H' \cdot W'\cdot R \cdot S \cdot D_1 \cdot D_2+H' \cdot W' \cdot N \cdot D_2$. The Tucker ranks $D_1$, $D_2$ are the main hyper-parameters that decide the overall complexity. Compared with the original convolution layer, the parameters and FLOPs reduction ratio $\gamma_P, \gamma_F$ are given by:
\begin{small}
\begin{align}
\gamma_P=&\frac{C \cdot N \cdot R \cdot S}{C \cdot D_1+R \cdot S \cdot D_1 \cdot D_2+N \cdot D_2},\\
\gamma_F=&\frac{H' \cdot W' \cdot R \cdot S \cdot C \cdot N}{H \cdot W \cdot C \cdot D_1+H' \cdot W' \cdot D_2 \cdot (R \cdot S \cdot D_1 + N)}.
\end{align}
\end{small}


\section{Hardware-aware Tucker Decomposition}
\label{sec:tucker-algorithm}

\subsection{Training Tucker-format Models with ADMM}
\begin{table}[b]
\caption{Accuracy comparison between directly training and our ADMM-based compression for ResNet-20 on CIFAR-10.}
    \footnotesize
    \centering
    \begin{tabular}{c|c|c}
    \toprule
     \textbf{Method}    &  \textbf{Top-1 (\%)} & \textbf{FLOPs$\downarrow$}\\\midrule
     Baseline & 91.25  & N/A\\ 
     Direct Compression & 87.41  & 60\%\\
     ADMM-based & 91.02 & 60\%\\\bottomrule
    \end{tabular}
    \label{tab:limitation_direct}
\end{table}
\textbf{Limitation of direct training.} The capacity of the Tucker-format model is lower and its depth becomes larger than the original model, thus it is more sensitive to hyper-parameters settings. As a result, directly training Tucker-format models from scratch suffers from significant performance degradation (Table \ref{tab:limitation_direct}). Another way to obtain a Tucker-format model is to decompose a pre-trained model, and then retrain. In this case, the pre-trained model is generally full rank, and decomposing it to Tucker format leads to a non-negligible approximation error. Thereby, the accuracy cannot be recovered in the Tucker-format model even after a long retraining.

Inspired by the previous optimization work \cite{yinadmm2021}, we propose an ADMM-based training technique to obtain high-accuracy 
Tucker-format models by gradually imposing low-Tucker-rank properties onto the original models during training process. The objective for compressing a baseline model using optimization-based training is given as
\begin{equation}
\label{eq:obj}
\min_{\bmmc{W}}~\ell(\bmmc{K}), ~~~\textrm{s.t.}~\mathrm{rank}(\bmmc{K})\leq [D_1^*, D_2^*],
\end{equation}
where $\ell$ is the loss function of the DNN model, $\mathrm{rank}(\cdot)$ returns the latent Tucker ranks of the kernel tensor $\bmmc{K}$, and $D_1^*, D_2^*$ are the desired Tucker ranks. 

Since $\mathrm{rank}(\cdot)$ is non-differentiable, the above training objective Eq. (\ref{eq:obj}) cannot be solved by a regular optimizer. Fortunately, Alternating Direction Method of Multipliers (ADMM) algorithms \cite{boyd2011distributed} can efficiently solve this non-convex problem via alternatively solving two simple sub-problems. As a result, the training is converted to iterative optimization.

First, with scaled augmented Lagrangian form, the original problem Eq. (\ref{eq:obj}) can be rephrased as
\begin{equation}\label{eq:lagrangian}
\min_{\bmmc{K},\widehat{\bmmc{K}}\in\mathcal{Q}}\max_{\bmmc{M}}~\ell(\bmmc{K}) + \frac{\rho}{2}\|\bmmc{K}-\widehat{\bmmc{K}}+\bmmc{M}\|_F^2 -\frac{\rho}{2}\|\bmmc{M}\|_F^2,
\end{equation}
where $\widehat{\bmmc{K}}$ is an introduced variable whose shape is identical to $\bmmc{K}$, $\mathcal{Q}$ is the set where kernel tensors satisfy the target constraints, i.e., $\mathcal{Q}=\{\widehat{\bmmc{K}}|\mathrm{rank}(\widehat{\bmmc{K}})\leq [D_1^*, D_2^*]\}$, $\bmmc{M}$ is the dual multiplier, and $\rho$ is the penalty coefficient.

Then, with ADMM, the alternate problem Eq. (\ref{eq:lagrangian}) can be split into two sub-problems which can be independently solved, and all the variables are iteratively updated in the training process.

\textbf{$\bmmc{K}$-update.} The first optimization step is to update the original kernel tensor, which is to solve
\begin{equation}
\min_{\bmmc{K}}~\ell({\bmmc{K}})+\frac{\rho}{2}\|\bmmc{K}-\widehat{\bmmc{K}}+\bmmc{M}\|_F^2.
\end{equation}
The above objective is essentially minimizing the original loss function with $L_2$-regularization. Hence, $\bmmc{K}$ can be updated via standard mini-batch SGD as
\begin{equation}
    \bmmc{K}\leftarrow\bmmc{K}- \alpha\left(\frac{\partial\ell(\bmmc{K})}{\partial\bmmc{K}}+\rho(\bmmc{K}-\widehat{\bmmc{K}}+\bmmc{M})\right),
\end{equation}
where $\frac{\partial\ell(\bmmc{K})}{\partial\bmmc{K}}$ is the gradient in the back-propagation and $\alpha$ is the learning rate.

\textbf{$\widehat{\bmmc{K}}$-update.} The following optimization step is to let $\widehat{\bmmc{K}}$ satisfy the desired Tucker ranks, i.e.,
\begin{equation}
\min_{\widehat{\bmmc{K}}}~\|\bmmc{K}-\widehat{\bmmc{K}}+\bmmc{M}\|_F^2, ~~~\textrm{s.t.}~\widehat{\bmmc{K}}\in\mathcal{Q}.
\end{equation}
According to \cite{boyd2011distributed}, this sub-problem can result in an analytical solution by directly projecting $\bmmc{K}+\bmmc{M}$ to set $\mathcal{Q}$. Hence, $\widehat{\bmmc{K}}$ can be explicitly updated via
\begin{equation}
\widehat{\bmmc{K}}\leftarrow \textbf{proj}(\bmmc{K}+\bmmc{M}),
\end{equation}
where \textbf{proj} is the truncated-HOSVD that truncates the smallest singular values of mode-1 and mode-2 matricization to satisfy the rank constraints. To be specific, suppose $\bmmc{T}=\bmmc{K}+\bmmc{M}\in\mathbb{R}^{C\times N\times R\times S}$, thus the mode-1 and mode-2 matricization of $\bmmc{T}$ is $\bm{T}_{(1)}\in\mathbb{R}^{C\times NRS}$ and $\bm{T}_{(2)}\in\mathbb{R}^{N\times CRS}$, respectively. Performing matrix SVD on $\bm{T}_{(1)}$ and $\bm{T}_{(2)}$, we have $\bm{U}_1\bm{\Sigma}_1\bm{V}_1=\bm{T}_{(1)}$ and $\bm{U}_2\bm{\Sigma}_2\bm{V}_2=\bm{T}_{(2)}$, where $\bm{\Sigma}_1$ and $\bm{\Sigma}_2$ are diagonal matrices with singular values in descending orders. To project $\bmmc{T}$ with target Tucker ranks, truncating smallest singular values in $\bm{\Sigma}_1$ and $\bm{\Sigma}_2$, with TKD we can obtain $\widehat{\bm{U}}_1\in\mathbb{R}^{C\times D_1^*}$,  $\widehat{\bm{U}}_2\in\mathbb{R}^{C\times D_2^*}$, and $\widehat{\bmmc{C}}\in\mathbb{R}^{C\times N\times D_1^*\times D_2^*}$. Correspondingly, we can use Eq. \ref{eq:tkd2} to recover back to the projected tensor $\widehat{\bmmc{K}}$ with the truncated $\widehat{\bm{U}}_1$, $\widehat{\bm{U}}_2$ and $\widehat{\bmmc{C}}$.

\textbf{$\bmmc{M}$-update.} The update step for dual multiplier $\bmmc{M}$ can be considered as the optimization for the dual problem (the maximizing problem described in Eq. \ref{eq:lagrangian}) by gradient descent with fixed step size $\frac{1}{\rho}$, i.e.,
$\bmmc{M}\leftarrow\bmmc{M}+\bmmc{K}-\widehat{\bmmc{K}}$.

In summary, the above three updates are alternatively performed during the optimization-incorporated training.

\subsection{Plug-in Hardware-Aware Constraints}

Even though one can obtain a highly accurate Tucker-format compressed model using the introduced optimization-based training technique in the previous subsection, it is still challenging to get the desired on-device speedup. This is because \underline{(1)} the decomposed model architecture is determined by the target Tucker ranks in Eq. (\ref{eq:obj}), which are set empirically. Thus, it difficult to fit all practical hardware devices; \underline{(2)} the compressed Tucker-format models are not well supported by existing convolution software (e.g., cuDNN) and deep learning framework (e.g., PyTorch), thereby leading to undesired latency in terms of theoretical FLOPs reduction. 

To address this challenge, 
we reformulate the optimization objective Eq. (\ref{eq:obj}) such that the rank setting with practical GPU latency can be considered as plug-in constraints, i.e.,
\begin{equation}\label{eq:plug-in}
\min_{\bmmc{W}}~\ell(\bmmc{K}), ~~~\textrm{s.t.}~\mathrm{rank}(\bmmc{K})\leq \mathcal{P}_{\textrm{device}}.
\end{equation}
Specifically, we first generate a benchmark with GPU performance for our designed kernel. Then, we find the proper Tucker ranks that can obtain the best latency and satisfy the overall compression budget. Correspondingly, we plug these rank settings as $\mathcal{P}_{\textrm{device}}$ in the above Eq. (\ref{eq:plug-in}). 

We will introduce \textsf{TDC} in detail in Section \ref{sec:co-design} after we present our design for efficient Tucker-format convolution.

\section{Our Convolution Kernel Design}
\label{sec:micor-kernel}


In this section, we first discuss the limitations of TVM’s convolution scheme and then propose our convolution scheme with a performance analysis. 

\subsection{Discussion on TVM's Convolution Scheme}
\label{sec:tvm-limit}
TVM follows direct convolution
to design its scheme \cite{chen2018tvm}. Specifically, the threads in each thread block are responsible for computing output positions with continuous coordinates on the same panel. Thus, the threads responsible for nearby positions have some overlapping from the perspective of the input tensor. Moreover, all threads in the same thread block require the same kernel weight elements. Each thread block keeps two shared memory buffers: one is for the input tensor and the other one is for the weight tensor. After loading the input to the shared buffer, each thread starts its own computation. Listing 1 shows its pseudo code. 

\textbf{Limitations of TVM's convolution scheme.} 
First, TVM introduces two thread synchronizations because its scheme uses the shared memory to achieve data reuse from both kernel tensor and input tensor (lines 1 and 2) at each iteration of C (input channel). 
Second, TVM's scheme splits the workloads on height and width dimensions for the threads. Hence, the tiles of input tensor are different for each thread. Accordingly, there exists overlapping of the tensor tile for each thread. To load all the required tiles of input tensor to the shared memory will potentially limit the achievable GPU occupancy. Moreover, since the tiles of input tensor are different across threads, they cannot load all tiles across C at one time and hence need to traverse the C loop with additional thread synchronizations {(line 6)}. Third, TVM's scheme does not split workloads across the input channels. 
However, convolutions in Tucker format have relatively small ranks for input/output channels, resulting in a smaller workload. Thus, not splitting across input channels leads to the under-utilization of resources.
\begin{lstlisting}[style=cuda, caption=\textcolor{black}{Pseudocode of TVM convolution design.},captionpos=b, label=lst:tvm,
numberstyle=\tiny\color{red}, basicstyle=\scriptsize, linewidth=\textwidth]
//Input: Input tensor (*@ $\mathcal{X}$ @*), Conv kernel (*@ $\mathcal{K}$ @*)
//Output: Output tensor (*@ $\mathcal{Y}$ @*)
shared float shared_input[]
shared float shared_kernel[]
float local_input[]
float local_kernel[]
float local_compute[]
for c = 0 to C: (*@ \label{alg1:cloop}  @*)
  threads_synchronization()
  load input to shared_input
  load kernel to shared_kernel
  threads_synchronization()
  load local_input
  load local_kernel
  for n =0 to N:
     local_compute
write_out
\end{lstlisting}
\vspace{-8mm}

\subsection{Proposed Tucker-format Convolutions}
Our proposed scheme for Tucker-format convolutions is also a direct-convolution-based scheme. In our design, we first tile the input tensor across its height (H), width (W), and input channel (C). 
There are ${\frac{H}{TH}\times\frac{W}{TW}\times \frac{C}{TC}}$ thread blocks in total, and each tile with the size of ${TH\times TW}$ is mapped to a thread block. 
We then assign every thread block $N$ threads (where $N$ is the number of output channels) and map each thread to one output channel.
After that, each thread block will load a cube with the size of ${(TH+R-1)\times(TW+S-1)\times TC}$ of the input tensor to its shared memory at the very beginning. 
After loading the tile of the input tensor to the shared memory, each thread will compute its contribution to the input tensor and save the intermediate result to a temporary array (in register) with the size of ${TH \times TW}$. In summary, each thread performs the following computation.
\begin{small}
\begin{equation*}\label{thread-work}
    \mathcal{Y}(n,th,tw)=\\
\sum_{c=0}^{TC-1}\sum_{r=0}^{R-1}\sum_{s=0}^{S-1}\mathcal{I}(c,th-r,tw-s)\times\mathcal{K}(n,c,r,s)
\end{equation*}
\end{small}

Finally, after finishing the computation, each thread will write the final output result from its temporary array to its mapped output channel. Hence, the input tensor is reused across all the output channels in our design. In other words, 
all the threads inside the thread block need the same tile of the input tensor.
Note that our proposed scheme splits the workload across the input channels, which not only provides a 
higher degree of parallelism but also enables us to load the input tensor at once. This can help avoid thread synchronizations which occur in the TVM's scheme. 
Furthermore, since each thread is responsible for a different output channel, there is no atomic writing required inside a thread block. 

Still, this scheme has a limitation---although splitting the workload across all output channels helps us get a fully reused input tensor and avoids thread synchronizations,
the data volume in the kernel is high, and the kernel data loaded for each thread is not usable for other threads. Thus, the data reuse rate 
in this scheme is lower than TVM's scheme. 

To mitigate this issue, we use the format of CRSN for the kernel tensor to reduce the time overhead of loading the kernel data.
This is because by using the CRSN format, the kernel tensor loading will be fully coalesced, which can reduce the time overhead. Note that the kernel tensor format conversion can be completely done offline once, 
which will not affect the inference performance. More details about our proposed convolution scheme can be found in Listing 2.

\begin{lstlisting}[style=cuda, caption=Pseudocode of our core convolution design.,captionpos=b, label=lst:alg,
numberstyle=\tiny\color{red}, basicstyle=\scriptsize, linewidth=\textwidth]
//Input: Input tensor (*@ $\mathcal{X}$ @*), Conv kernel (*@ $\mathcal{K}$ @*)
//Output: Output tensor (*@ $\mathcal{Y}$ @*)
shared input_tile[TC][(TH+R-1)*(TW+S-1)]
float temp_result[TH][TW], kernel[R][S]
unsigned int tile_tc_id = blockId/(H/TH * W/TW)
unsigned int tile_id = blockId%(H/TH * W/TW)
unsigned int tile_h_id = tile_id/(W/TW)
unsigned int tile_w_id = tile_id%(W/TW)
unsigned int output_n = threadIdx.x (*@ \label{alg1:oid}  @*)
//copy tiled input tensor from global to shared
copy(input_tile,(*@ $\mathcal{X}$ @*))
syncthreads() //synchronize all threads in a thread block (*@ \label{alg1:sync}  @*)
for c = 0 to TC: (*@ \label{alg2:cloop}  @*)
  copy(kernel,(*@ $\mathcal{K}$ @*),n,c+tile_tc_id*TC)  (*@ \label{alg1:kcopy}  @*)
  for (v,h,w) in (input_tile): (*@ \label{alg1:eloop}  @*)
    for r = 0 to R
      for s = 0 to S
        y_out = h - r
        x_out = w - s
        if y_out<0 or x_out< 0 or y_out>TH or x_out>TW:
            continue
        result = v * kernel[r][s] (*@ \label{alg1:cm1}  @*)
        temp_result[y_out*TW+x_out] += result (*@ \label{alg1:cm2}  @*)
// Write the output back to memory
for th to TH: (*@ \label{alg1:iterout}  @*)
  for tw to TW:
    y = tile_id/(W/TW)*TH+th
    x = tile_id%(W/TW)*TW+tw
    atomicAdd(Y[H*W*N+y*W*N+x*N+n],temp_result[th*TW+tw])
\end{lstlisting}
\vspace{-8mm}

\subsection{Computation latency analysis}
In our convolution kernel for Tucker core tensors, tiling plays an important role to determine the data movement volume and computation resource utilization. In the following discussion, we will model computation and memory latency based on a given tiling size and convolution shape. Based on these models, we will propose our tiling size selection.

From the perspective of resource utilization: the total number of thread blocks can be denoted as
$\small
\text{num\_blks} = \frac{H}{TH} \times \frac{W}{TW} \times  \frac{C}{TC}$.
Since each thread block has $N$ threads, the total number of threads in our core convolution kernel is
$\small \text{Num\_ths} = \text{Num\_blks} \times N$.
The number of FLOPs for each thread block is
$\small \text{flops\_blk} = 2\times (TH+R-1)\times (TW+S-1)\\\times TC\times N \times R \times S$.
The peak performance for each thread block is
$\small \text{blk\_peak} = \text {GPU\_peak} \times \frac{N}{\text{GPU\_ths}}$.
Thus, the computation latency for each thread block can be estimated as
$\small \text{comp\_latency\_blk} = \frac{\text{FLOPs\_blk}}{\text{Blk\_peak}}$.
If we combine all the equations above, we can derive the computation latency for a thread block as $\text{comp\_latency\_blk} = $
\begin{equation*} 
\small
\frac{ 2\cdot (TH+R-1) \cdot (TW+S-1)\cdot TC\cdot \text {GPU\_ths} \cdot R \cdot S}{ \text {GPU\_peak} }
\end{equation*}

When launching a GPU kernel, if the number of threads requested is greater than the total threads that a GPU can provide, the computation will be divided into multiple waves, and the kernel can be completed only after the last wave finishes.
We formulate the number of GPU waves as
\begin{equation} \label{waves}
\small
\begin{split}
\text{comp\_waves} & = \left \lceil \frac{\text{Num\_ths}}{\text{GPU\_ths} \times \text{Occupancy}}\right \rceil\\
& =\left \lceil\frac{\frac{H}{TH} \times \frac{W}{TW} \times  \frac{C}{TC} \times N}{ \text{GPU\_ths} \times \text{Occupancy} } \right \rceil.
\end{split}
\end{equation}
Eq. (\ref{waves}) illustrates that the total number of GPU waves is determined by the convolution problem shape ($H$, $W$, $C$, $N$), tiling sizes ($TH$, $TW$, $TC$) and the GPU hardware metrics ($\text{GPU\_ths}$). Note that the occupancy can be estimated by the hardware metrics such as shared memory size, register file size along with the given tiling sizes ($TH$, $TW$, $TC$); or simply put we can obtain it by querying via the NVCC compiler to get the exact occupancy for specific tiling sizes.

The computation latency for each wave is equal to the computation latency of a thread block as all the thread blocks in a wave have the same FLOPs for a dense convolution. Hence, the total computation latency for the kernel is
\begin{equation} 
\text{comp\_latency} = \text{comp\_waves} \cdot \text{comp\_latency\_blk}.
\end{equation}

When only considering the computation latency, 
an optimized tiling configuration ($TH$, $TW$, $TC$) for a given convolution shape ($H$, $W$, $C$, $N$) is to minimize ${Comp\_latency}$ under the constraint of ${TH \leq H, TW \leq W, TC \leq C}$. 

\subsection{Memory latency analysis}
On the GPU, besides computation latency, memory latency also plays an important role in the overall latency. Different tiling sizes affect not only the computation latency but also the data movement, which further influences the memory latency. Different tiling selections will affect the data movement from different GPU memory hierarchies such as shared memory, register, and global memory. Since global memory is the slowest memory, we only analyze the data movement from the perspective of global memory. For a given tiling size ($TH$, $TW$, $TC$) and convolution shape ($H$, $W$, $C$, $N$), the data movement volume of convolution kernel can be denoted as
\begin{equation} \label{data volume for kernel}
\begin{split}
volume_k = \left \lceil \frac{H}{TH} \right \rceil \times \left \lceil \frac{W}{TW} \right \rceil \times C \times N.
\end{split}
\end{equation}
The input tensor volume can be denoted as
\begin{equation} \label{data volume for input}
\begin{split}
volume_x = \left \lceil \frac{H}{TH} \right \rceil \times \left \lceil \frac{W}{TW}\right \rceil \times C \times \\(TH+R-1)\times (TW+S-1).
\end{split}
\end{equation}
The output tensor volume can be denoted as
\begin{equation} \label{data volume for output}
\begin{split}
volume_y = H \times W \times N \times \frac{C}{TC}.
\end{split}
\end{equation}
Thus, the total data-movement volume can be denoted as
\begin{equation} \label{total volume}
\begin{split}
volume\_total = volume_x + volume_k + volume_y.
\end{split}
\end{equation}
Based on Eq. (\ref{data volume for kernel}), (\ref{data volume for input}), (\ref{data volume for output}), we can also find an optimized tiling option to minimize the overall data movement latency.
%
\subsection{Analytical model for tiling selection}
The previous analyses show that the tiling option has a significant effect on both computation latency (${comp\_latency}$) and memory latency (${memory\_latency}$). An ideal tiling option will minimize both latencies. However, the best tiling option for the computation latency may not be the best option for the memory latency, vice versa. Hence, the best tiling option should minimize the overall latency. 

We note that a previous work \cite{park2016faster} demonstrates that dense convolution is likely to be compute bound. Thus, in our tiling analytical model, we put the computation latency in prior to the memory latency. 
Specifically, our tiling analytical model first computes the ${comp\_latency}$ for all the tiling size options. Based on the results, we sort the tiling configurations in increasing order. Then, we select the top (5\%, 15\%) of tiling configurations (A100, 2080Ti) as our candidates. For those selected candidates, we pick the one which has the minimum  ${memory\_latency}$ as our final tiling selection. Our tiling analytical model allows to generate high-performance convolution code on GPUs for a given shape without going through a costly parameter tuning process required by TVM. 

In addition to our analytical modeling, we also provide an auto-tuning script for core convolutions as part of our framework. The auto-tuning process is based on an exhaustive search of tiling size. For a give core convolution shape, we run all tiling options (the number of tiling options is ${H\times W\times C}$) and obtain the best one based on their actual latency. The exhaustive search guarantees to produce the best tiling option for a given convolution shape. However, it requires a high time overhead of offline tuning like TVM. The code generated by our analytical model achieves a slightly worse performance compared with exhaustive search (called ``oracle''), i.e., $\sim$25\% performance drop on both A100 and 2080Ti machines, while it is still 1.5$\times$ faster than TVM on average.
The performance comparison of the code generated with ``oracle'' and ``model'' approaches will be shown later. 

Our analytical modeling provides the user an option to generate a fast Tucker-friendly convolution kernel for quick deployment; while if the user looks for extreme performance and does not care about the costly offline tuning process, using the exhaustive search is recommended.

\section{Our Proposed Co-design Framework}
\label{sec:co-design}

In our Tucker decomposition, an original convolution kernel will be two 1$\times$1 convolutions and one core convolution. The filter of our core convolution has the same height ($R$) and width ($S$) compared with the original convolution. The input channels ($D_1$) and output channels ($D_2$), i.e., the ranks for the core convolution ($C,N$), for our core convolution are smaller than the original convolution. The smaller ($D_1,D_2$), the more FLOPs we can reduce. Theoretically, the ranks ($D_1,D_2$) for our core convolution can be reduced to any size. Greedily, we can reduce them until (1,1). However, if the ranks are too small, it causes two problems: \underline{(1)} 
It would cause a significant drop on model accuracy (hard to recover).
\underline{(2)} It would decrease the potential for pruning the next convolution layers. 

\begin{figure}[t]
    \centering
    \includegraphics[width=0.8\linewidth]{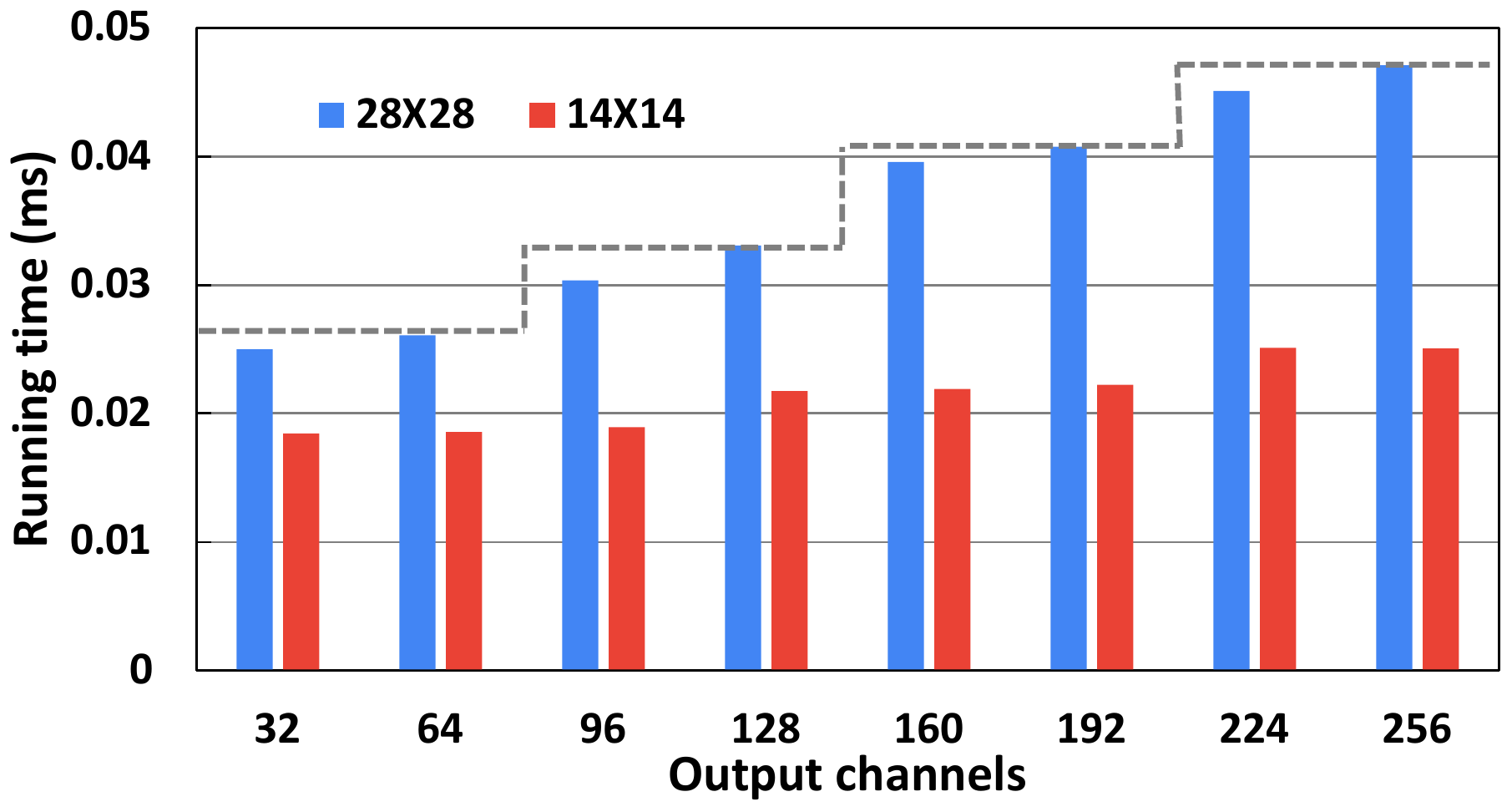}
    \caption{Runtime of convolutions with different numbers of output channels (the number input channels is fixed to 64).}
    \vspace{-2mm}
    \label{fig:performance trend}
\end{figure}

The fundamental idea of our proposed co-design framework is only considering the reduced ranks that can provide a performance/runtime benefit. In Section \ref{sec:micor-kernel}, we have demonstrated that different convolution shapes may end up to with the same computation latency by adjusting tiling size. Figure \ref{fig:performance trend} shows the overall latency changing trend for two convolutions with their ${C,H,W}$ fixed and ${N}$ changes from 32 to 256 on 2080Ti machine. The blue line has ${C=64,H=28,W=28}$ and the red line has ${C=64,H=14,W=14}$. Both of their running time trend lines are in a monotonic staircase  when the  output channels (${N}$) increase. It shows the fact that the overall latency may remain the same when the FLOPs changes. The most important reason is that an optimized tiling strategy can increase the parallelism level for a larger problem to reduce the overall latency. Intuitively, we can employ this to guide our decomposition and training process to avoid the ``over rank reduction'' situation.

For a given trained CNN, users need to first carefully determine a decomposition budget ${B}$ (i.e., the target FLOPs reduction ratio). \textcolor{black}{An aggressive budget may lead to accuracy drop, so we recommend to choose ${B}$ based on state-of-the-art FLOPs reduction ratios.
While in the scenario where there is no FLOPS reduction ratio for reference, we recommend to start the budget from 10\%; we can further increase the budget as long as the current budget meets the accuracy requirement after training for some epochs.
}

\begin{figure}[b!]
    \centering
    \includegraphics[width=0.7\linewidth]{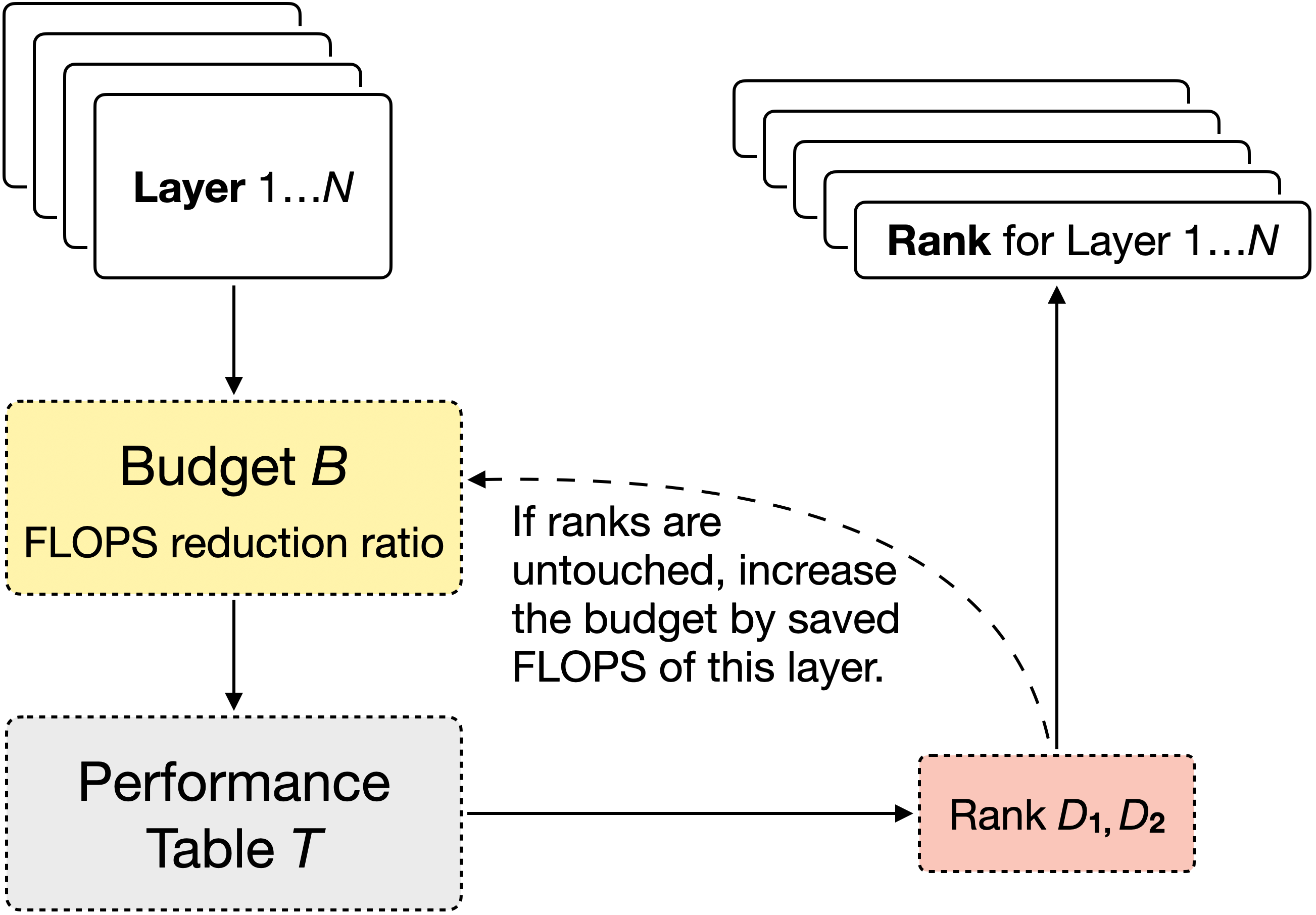}
    \caption{\textcolor{black}{Workflow of proposed hardware-ware rank selection.}}
    \vspace{-2mm}
    \label{fig:codesign flow chart}
\end{figure}

After that, we select our decomposition ranks for retraining based on ${B}$. 
The entire process of our rank selection is shown in Figure \ref{fig:codesign flow chart}. 
Specifically, the first step 
is to generate the micro-kernel CUDA code for all possible decomposed convolutions for each original convolution in the model by applying their optimized tiling size (as described in Section \ref{sec:micor-kernel}), and empirically run them to get a performance table ${T}$.
Theoretically, for a convolution layer with shape ${(C,N,H,W)}$, there exist ${C\times N}$ decomposition candidates because the decomposed input channel can be any number from $\{1,2,3...,C-1\}$ and the output channel can be any number from $\{1,2,3...,N-1\}$. 
However, reducing the input and output channels by one at a time is unnecessary for two reasons: (1) it reduces FLOPs by very small amounts, and (2) it may cause idle threads in some convolution schemes 
as a GPU warp performs as a group of 32 threads.
To this end, 
we reduce the input channels and output channels by 32 each time. Hence, there exist ${\frac{C}{32} \times \frac{N}{32}}$ decomposition candidates.

For each convolution layer, assuming the number of input channels is C and the number of output channels is N. With a budget of ${B}$, we choose the $(D_1,D_2)$ such that the overall FLOPs in the Tucker-format model is less than the budget, i.e., $\mathcal{P}(D_1,D_2) \lessapprox B$, where $D_1$ and $D_2$ are the Tucker ranks that determine the input channel and output channel of the core convolution. Note that since Tucker decomposition allows to decrease both ${C}$ and ${N}$, 
there will be multiple $(D_1,D_2)$ candidates.
Thus, we further look up the performance table ${T}$ to select $(D_1,D_2)$ based on its performance/latency.

\textcolor{black}{
Considering that the Tucker decomposition brings two more ${1\times 1}$ convolutional layers, and the extra two ${1\times 1}$ of kernel launch time may even cause the performance worse than the original convolutional layer, we use a threshold $\theta$ to determine whether $(D_1,D_2)$ should be selected. Specifically, assuming $(D_1,D_2)$ has a latency of ${t_1}$ and the original $(C,N)$ convolution layer has a latency of ${t_2}$, 
we will not apply Tucker decomposition on this layer if ${t_1 \geq (1 - \theta) \times t_2}$. For simplicity, we choose $\theta = 15\%$ in our experiments.
}
Furthermore, if this convolution layer is unchanged, it will save ${C \times R \times S \times H \times W \times N \times B }$ FLOPs in budget which can be further utilized; in other words, we can increase ${B}$ by the saved FLOPs reduction to the remaining convolution layers.

Finally, after going through all convolution layers, we can obtain $(D_1^*,D_2^*)$ as our target ranks for Tucker decomposition and 
plug them into Equation (\ref{eq:plug-in}) to perform the ADMM-based training to fine-tune for several epochs, 
obtaining the high-performance model with the lowest latency.
Algorithm \ref{alg:co-design-training} shows the detail of our co-designed training process.

\begin{algorithm}[t]
\caption{\textsf{TDC}: Proposed co-design framework for hardware-aware Tucker decomposed training.}
\label{alg:co-design-training}
\footnotesize
\SetAlgoLined
\textbf{Input:} Pre-trained CNN kernel $\bmmc{K}$, decompose budget $B$, optimized benchmark table $T$, training epochs $E$;\\
\textbf{Output:} {Hardware-aware Tucker decomposed kernels $\bm{U}_1,\bm{U}_2,\bmmc{C}$.}\\
$D_1^*,D_2^*=\max\{\argmin_{\mathcal{P}(D_1,D_2)\le B}~T(D_1, D_2)\}$; \textcolor{Green}{//Minimize the overall latency while maximize ranks under the budget}\\
Plug $D_1^*, D_2^*$ to Equation (\ref{eq:plug-in});\\
$\widehat{\bmmc{K}}\leftarrow \bmmc{K}, \bmmc{M}\leftarrow 0$;\\
\textcolor{Green}{//ADMM-incorporated training}\\
\While{$e<E$}{
    $\bmmc{K}\leftarrow\bmmc{K}- \alpha\left(\frac{\partial\ell(\bmmc{K})}{\partial\bmmc{K}}+\rho(\bmmc{K}-\widehat{\bmmc{K}}+\bmmc{M})\right)$;\\
    $\widehat{\bmmc{K}}\leftarrow \textbf{proj}(\bmmc{K}+\bmmc{M})$;\\
    $\bmmc{M}\leftarrow\bmmc{M}+\bmmc{K}-\widehat{\bmmc{K}}$;\\
}
$\bm{U}_1,\bm{U}_2,\bmmc{C}=\mathrm{TKD}(\bmmc{K})$; \textcolor{Green}{//Decompose to Tucker format}\\
Fine-tune $\bm{U}_1,\bm{U}_2,\bmmc{C}$ in the Tucker-format model.\\

\end{algorithm}

\section{Performance Evaluation}
\label{sec:evaluation}

\subsection{Experimental Setup}
We use two platforms for evaluation: (1) Nvidia GTX 2080 Ti (68 SMs, 11 GB) machine running Ubuntu 20.04 LTS and, and (2) Nvidia Ampere A100 (108 SMs, 80GB) machine running Ubuntu 18.04.4 LTS. We use CUDA 11.3.1 paired with cuDNN 8.2.0 on the 2080Ti platform and CUDA 11.6.1 paired with cuDNN 8.2.1 on the A100 platform. These platforms represent two different GPU architectures (Ampere and Turing), which are designed for two different user groups (consumer and enterprise). On both of the platforms, we use \textcolor{black}{TVM 0.7.0} paired with LLVM 10.0.0.
For cuDNN's convolution operations,  regarding the layerwise performance evaluation, we compare our micro-kernel with three different cuDNN convolution methods (i.e., GEMM-based convolution \texttt{IMPLICIT\_GEMM} \cite{implicit-gemm-conv}, WINOGRAD-based convolution \texttt{WINOGRAD} \cite{conv-algo}, and FFT-based convolution \texttt{FFT} \cite{conv-algo}) and TVM;
regarding the end-to-end performance evaluation, we fix the cuDNN convolution method as \texttt{IMPLICIT\_GEMM} \cite{implicit-gemm-conv} since it has the best performance compared with the other two cuDNN convolution methods 
on the 2080Ti machine. 
Moreover, in the experiment, we run 1,000 inferences with different inputs for each model and report the average time.

\begin{figure*}[!ht]
    \centering
    \vspace{-2mm}
    \includegraphics[width=.88\textwidth]{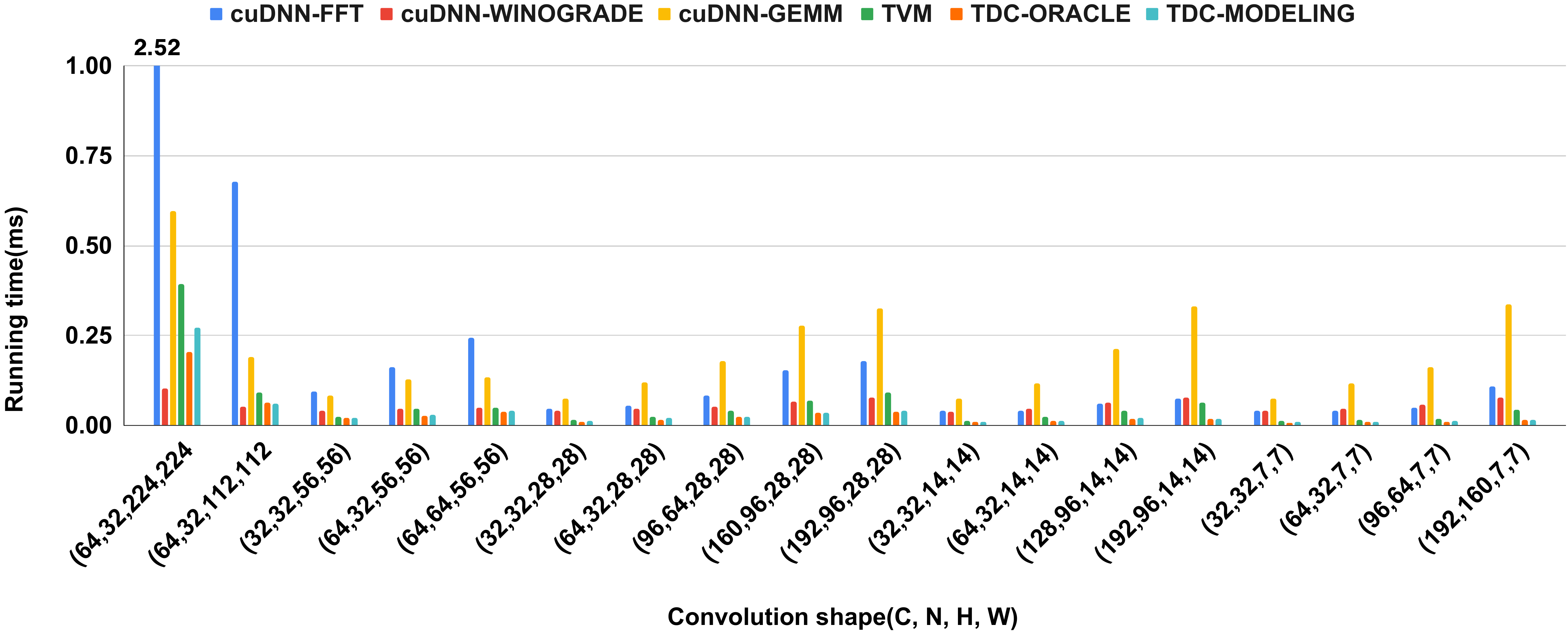}
    \caption{Performance comparison of our proposed kernel with TVM and cuDNN on all convolution shapes in the tested CNN models on A100.}
    \vspace{-2mm}
    \label{fig:a100 conv layer bencmark}
\end{figure*}

\begin{figure*}[!ht]
    \centering
    \includegraphics[width=.88\textwidth]{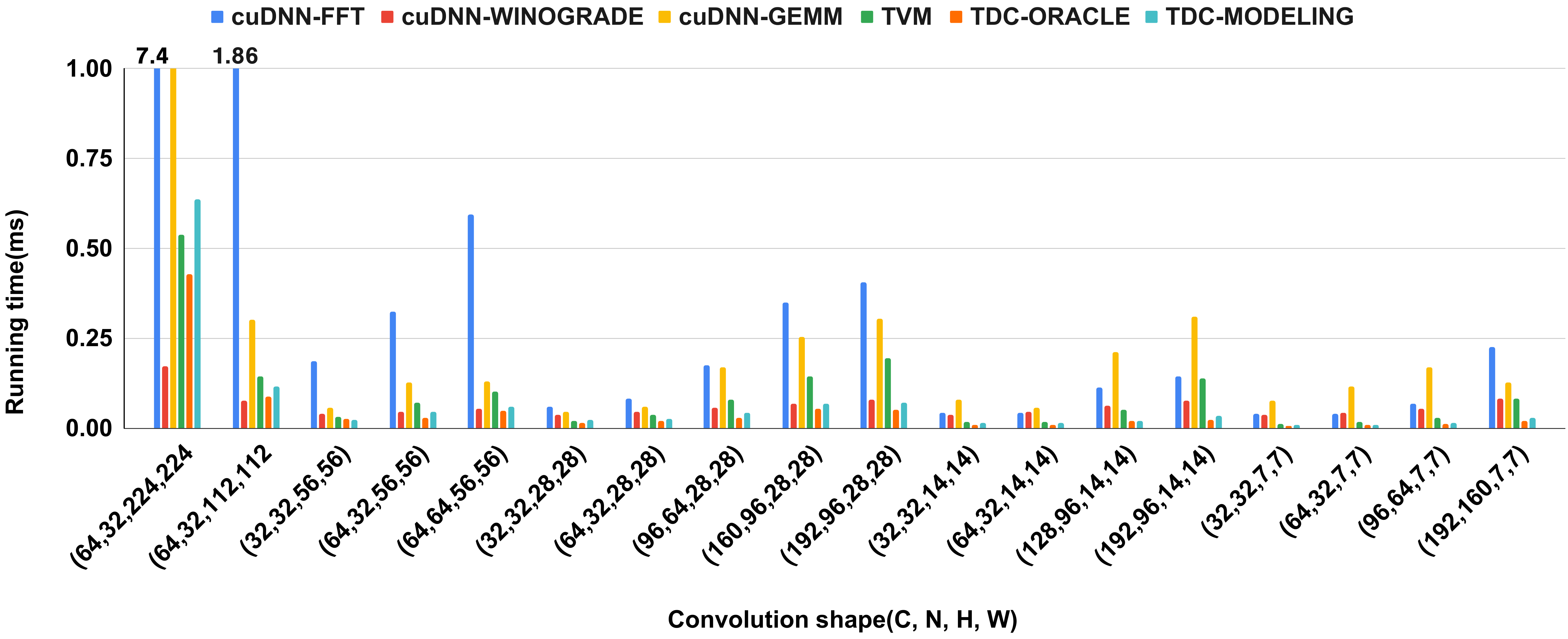}
    \caption{Performance comparison of our proposed kernel with TVM and cuDNN on all convolution shapes in the tested CNN models on 2080Ti.}
    \vspace{-4mm}
    \label{fig:2080Ti conv layer benchmark}
\end{figure*}

We choose five CNN models including one VGG \cite{simonyan2014very}, two DenseNets \cite{huang2017densely}, and two ResNets \cite{he2016deep} on large-scale dataset ImageNet \cite{deng2009imagenet}. 
These three types of networks represent the architecture of modern CNNs. However, modern CNNs also grow in width. For example, there are networks which are not deep but are wide such as GoogleNet \cite{szegedy2015going} and NasNet \cite{zoph2018learning}. In those CNNs, there are multiple convolutions computed at the same time at each convolution stage. We leave those wide CNNs to our future work due to two factors: (1) It is difficult to develop a scheme that minimizes the latency for multiple concurrent convolutions. (2) It is challenging to determine the ranks for the concurrent convolutions in the same stage.

\subsection{Evaluation on Model Accuracy}
\label{sec:eval-accuracy}

\begin{table}[t]
    \caption{Performance comparison between \textsf{TDC} and existing SOTA compression methods. MD denotes matrix decomposition. DN-1/-2 denote DenseNet-121/-201. Most existing tensor decomposition works do not report results for ResNet-50, VGG-16 and DenseNet on ImageNet.}
    \resizebox{.35\textwidth}{!}{
    \centering
    \footnotesize
    \begin{tabular}{cl|c|c|c}
\toprule
 \multicolumn{2}{c|}{\multirow{2}{*}{\makecell{\textbf{Model}}}} & \multirow{2}{*}{\makecell{\textbf{Compression}\\\textbf{Method}}} & \multirow{2}{*}{\makecell{\textbf{Top-1/Drop}\\\textbf{(\%)}}} & \multirow{2}{*}{\makecell{\textbf{FLOPs}$\downarrow$}}\\
& & &\\
\midrule
\multicolumn{1}{c}{\multirow{10}{*}{\rotatebox{90}{ResNet-18}} }
&\multicolumn{1}{|c|}{Original \cite{he2016deep}} & No compr. & 69.75/-0.00 & N/A\\
&\multicolumn{1}{|c|}{FPGM \cite{he2019filter}} & Pruning & 68.41/-1.34 & 42\%\\
&\multicolumn{1}{|c|}{DSA \cite{ning2020dsa}} & Pruning & 68.61/-1.14 & 40\%\\
&\multicolumn{1}{|c|}{SCOP \cite{tang2020scop}} & Pruning & 68.62/-1.13 & 45\%\\
&\multicolumn{1}{|c|}{TRP \cite{xu2020trp}} & MD & 65.51/-4.24 & 60\%\\
&\multicolumn{1}{|c|}{Stable \cite{phan2020stable}} & CPD & 69.06/-0.69 & 65\%\\
&\multicolumn{1}{|c|}{Opt. TT \cite{yinadmm2021}} & TTD & 69.29/-0.46 & 60\%\\
&\multicolumn{1}{|c|}{Std. TKD \cite{kim2015compression}} & TKD & 66.65/-3.10 & 60\%\\
&\multicolumn{1}{|c|}{MUSCO \cite{gusak2019automated}} & TKD & 69.28/-0.47 & 58\%\\
&\multicolumn{1}{|c|}{TDC} & TKD & 69.70/-0.05 & 63\%\\
\midrule
\multirow{6}{*}{\rotatebox{90}{ResNet-50}} 
&\multicolumn{1}{|c|}{Original \cite{he2016deep}} & No compr. & 76.13/-0.00 & N/A\\
&\multicolumn{1}{|c|}{FPGM \cite{he2019filter}} & Pruning & 75.59/-0.54 & 42\%\\
&\multicolumn{1}{|c|}{HRank \cite{lin2020hrank}} & Pruning & 74.98/-1.15 & 44\%\\
&\multicolumn{1}{|c|}{TDC} & TKD & 77.46/+1.33 & 40\%\\
&\multicolumn{1}{|c|}{Stable \cite{phan2020stable}} & CPD & 74.66/-1.47 & 60\%\\
&\multicolumn{1}{|c|}{TDC} & TKD & 76.42/+0.29 & 60\%\\
\midrule
\multirow{3}{*}{\rotatebox{90}{VGG-16}} 
&\multicolumn{1}{|c|}{Original \cite{he2016deep}} & No compr. & 71.59/-0.00 & N/A\\
& \multicolumn{1}{|c|}{CC \cite{li2021towards} } & MD & 68.81/-2.78 & 50\%\\
& \multicolumn{1}{|c|}{TDC} & TKD & 71.62/+0.03 & 80\%\\\midrule
\multirow{2}{*}{\rotatebox{90}{DN-1}}
&\multicolumn{1}{|c|}{Original \cite{he2016deep}} & No compr. & 74.43/-0.00 & N/A\\
&\multicolumn{1}{|c|}{TDC} & TKD & 76.33/+1.90 & 10\%\\\midrule
\multirow{2}{*}{\rotatebox{90}{DN-2}}
&\multicolumn{1}{|c|}{Original \cite{he2016deep}} & No compr. & 76.88/-0.00 & N/A\\
&\multicolumn{1}{|c|}{TDC} & TKD & 76.92/+0.04 & 10\%\\\bottomrule
    \end{tabular}
}
    \label{tab:accuracy}
\end{table}

We evaluate \textsf{TDC} and compare it with other state-of-the-art CNN model compression approaches including pruning based approaches such as FPGM \cite{he2019filter}, DSA \cite{ning2020dsa}, SCOP \cite{tang2020scop}, the matrix decomposition based approach such as TRP \cite{xu2020trp}, and other tensor decomposition approaches.
Table \ref{tab:accuracy} shows that the test accuracy and FLOPs reduction with different methods. 
For the budget $B$, we choose 65\%/60\% for ResNet-18/-50, since state-of-the-art works have achieved these FLOPs reductions; similarly, we choose 50\% for VGG-16 at the beginning, leading to a much higher accuracy than the state-of-the-art work CC \cite{li2021towards}, so we increase the budget to 65\% and further to 80\%; for DN-1/-2, due to no existing work to refer to FLOPs reduction ratio, we choose 10\% as the budget.
It is seen that our compression method provides the highest test accuracy even with the highest FLOPs reduction for all tested DNN models.
Specifically, our TKD-compressed models reduce the FLOPs by up to 63\% with only 0.05\% accuracy drop compared with the original non-compressed models on the tested CNNs. 
Note that except ResNet-18, the inference accuracy of our TKD-compressed models even has higher accuracy than that of the original non-compressed models, thanks to the proposed training algorithm. 

\textcolor{black}{
We also evaluate the impact of different target budgets on accuracy. We choose 65\%, 70\%, 75\%, and 80\% as our target budgets for ResNet-18. The achieved accuracies are 69.70\%, 67.86\%, 66.59\%, and 64.81\%, respectively, and the achieved FLOPs reductions are 66\%, 70\%, 76\%, and 80\%, respectively. It illustrates that aggressive target budgets lead to significant accuracy drops. Thus, we recommend users choose the budget $B$ based on existing work (e.g., 65\% in ResNet-18 from \cite{li2021towards}) or starting from 10\% if no prior knowledge. 
}

\subsection{Evaluation on Convolution Kernel Performance}
\label{sec:kernel-performance}
To demonstrate the effectiveness of our proposed convolution kernel, we compare the runtime of our kernel against the convolution kernel generated by TVM code generator and provided by cuDNN. 
We use the data type of float32 (single precision), the filter size of 3 (for core convolution), and the batch size of 1 (for inference).

Figure \ref{fig:a100 conv layer bencmark} and Figure \ref{fig:2080Ti conv layer benchmark} show the speedup of our scheme compared to TVM and cuDNN across all the core convolution shapes in the networks. On A100, our convolution with oracle/modeling approaches achieve a 5.38$\times$/4.91$\times$ speedup over cuDNN-FFT, 3.12$\times$/2.92$\times$ speedup over cuDNN-WINOGRAD, 8.95$\times$/8.63$\times$ speedup over cuDNN-GEMM, and 1.81$\times$/1.72$\times$ over TVM on average; on 2080Ti, our convolution with oracle/modeling approaches achieves 8.17$\times$/6.21$\times$ speedup over cuDNN-FFT, 2.75$\times$/2.12$\times$ speedup over cuDNN-WINOGRAD, 5.84$\times$/5.38$\times$ speedup over cuDNN-GEMM, and 2.35$\times$/1.81$\times$ over TVM on average.\par

From Figure \ref{fig:a100 conv layer bencmark} and Figure \ref{fig:2080Ti conv layer benchmark}, we can observe that there are two shapes (64, 32, 224, 224) and (64, 32, 112, 112) from VGG16 where our solution is actually slower than or similar to TVM and cuDNN. The main reason is that our scheme splits the workload across the output channels inside a thread block, while TVM splits the workload across height/width. Hence, for the convolution with large height and width, our scheme suffers a large kernel volume (referred to Equation (\ref{data volume for kernel})). However, we argue that this type of shape is becoming rare in modern CNN architecture designs. 
This is because the height (H) and width (W) of convolution layers decrease rapidly when the model goes deeper (mainly due to pooling layers), and hence the input channel (C) and output channel (N) increase. 
After an input goes into the model, after one or two layers, H/W decrease to a medium size, and the convolution shapes are likely to have very few large H/W but many medium H/C with large C/N.
For example, in the CNNs released in recent years such as DenseNet, ResNet, and GoogleNet, the largest H/W is 56 for all the convolution layers (except the first convolution layer for the input). VGG is still considered one of the modern CNNs, however, it is older than the other tested models with lower accuracy.

\subsection{Evaluation on End-to-End Inference Time}
To fully demonstrate that our proposed co-design framework is effective not only for core convolutions but also for the entire network, we present the end-to-end performance (involving all layers) on the tested CNN models. 
To better understand the performance improvement, we implement the five models using C++/CUDA code to eliminate the overhead of Python deep learning framework. For example, in PyTorch, padding operation is done on the CPU side which requires additional memory copy between device and host. 

\begin{figure}[t!]
    \centering
    \includegraphics[width=0.9\linewidth]{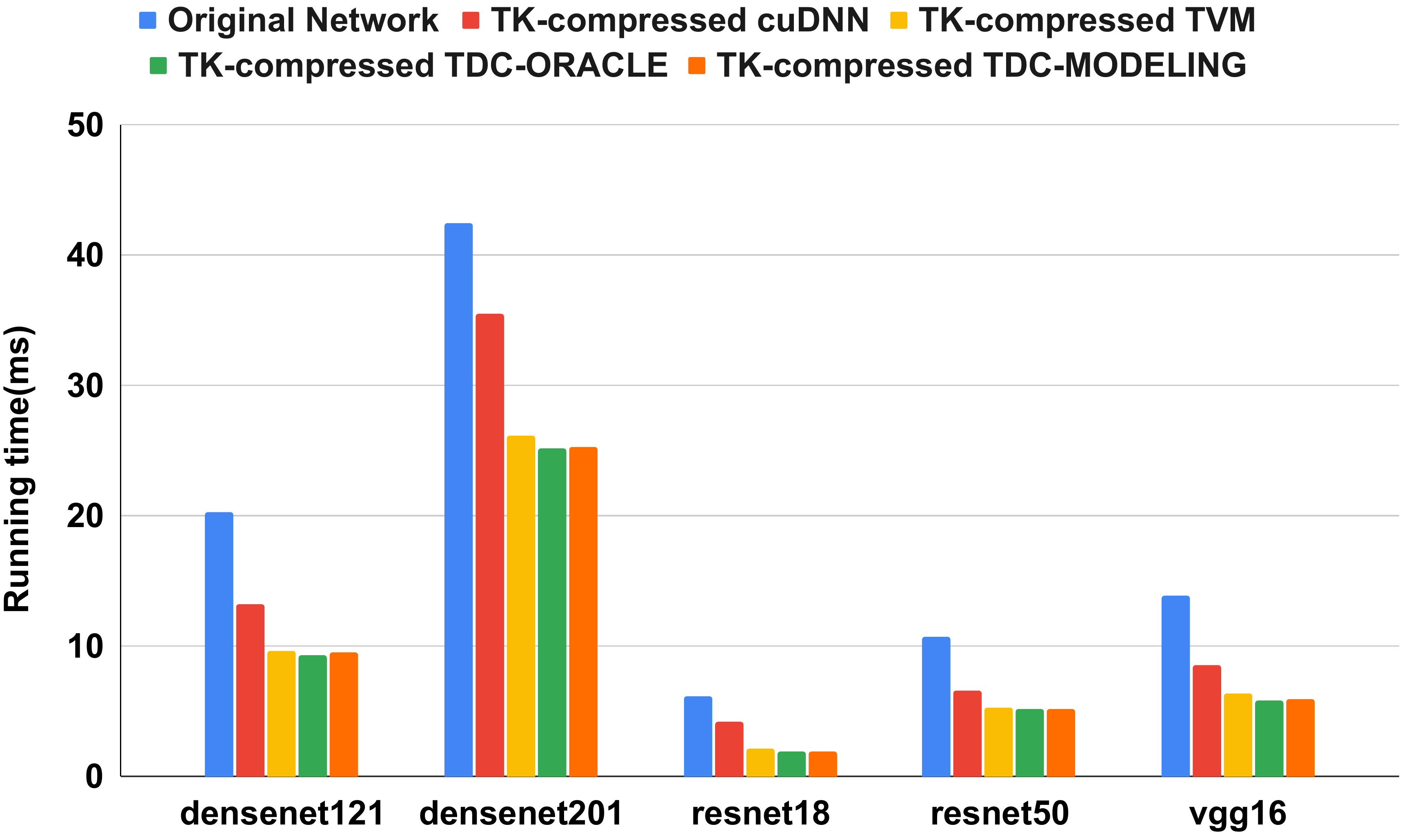}
    \caption{End-to-end evaluation of inference performance on A100.}
    \label{fig:A100-end2end}
    \vspace{-8mm}
\end{figure}
\par
\begin{figure}[t!]
    \centering
    \includegraphics[width=0.9\linewidth]{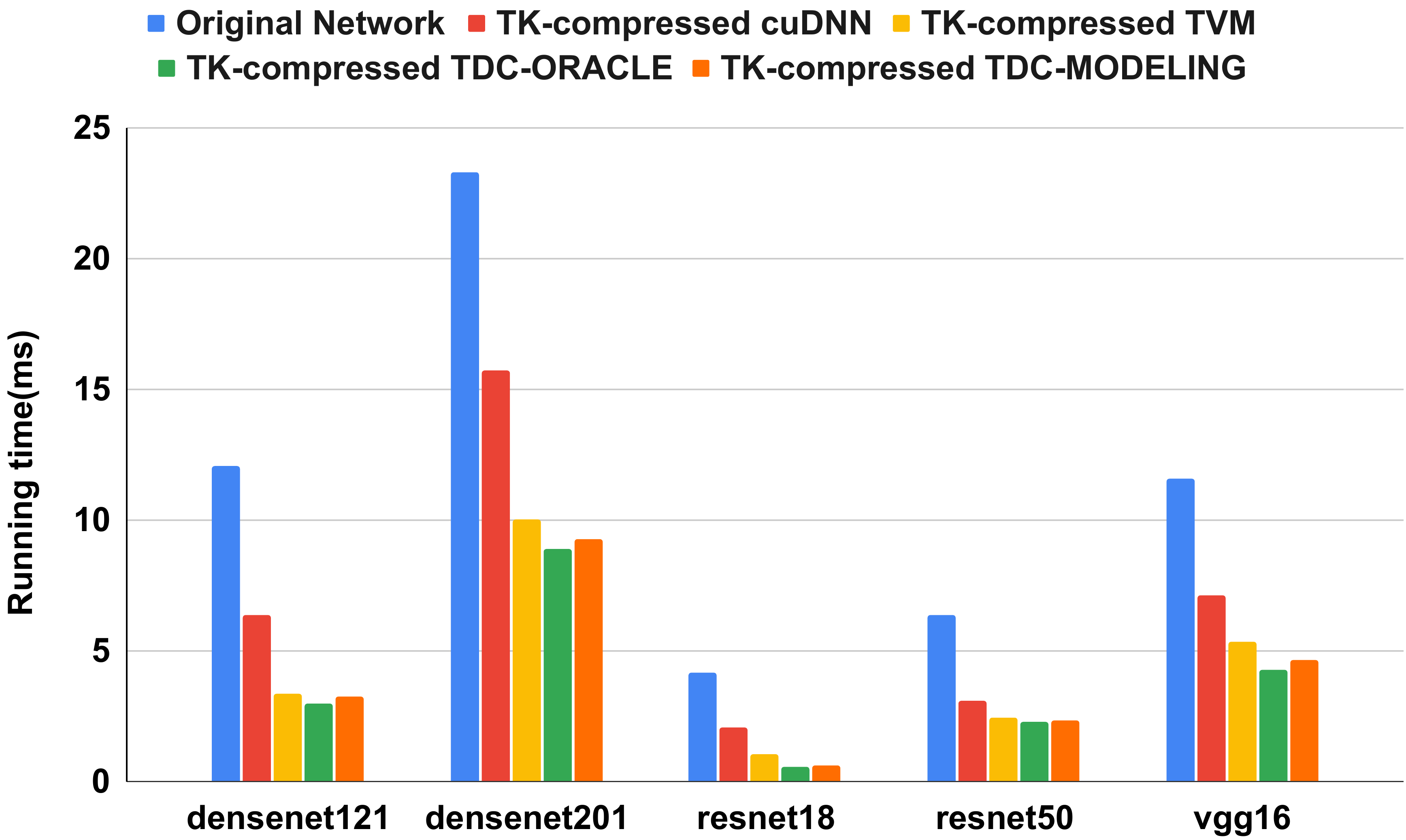}
    \caption{End-to-end evaluation of inference performance on 2080Ti.}
    \label{fig:2080ti-end2end}
    \vspace{-2mm}
\end{figure}

Figures \ref{fig:A100-end2end} and \ref{fig:2080ti-end2end} show the inference time comparison among original models, TKD-compressed models with cuDNN, and TKD-compressed models with our optimized convolution operations. 
Specifically, the blue bars show the inference time of the original models on A100 and 2080Ti. For the implementation of original models, we call the cuDNN library for all the layers including the convolution layers. 
The red bars represent the inference time of TKD-compressed models which also call the cuDNN library for all the layers. The yellow bars represent the inference time of TKD-compressed models using TVM generated kernel for the core convolutions. The green bars represent the inference time of TKD-compressed models using our optimized kernel for the core convolutions. 
Note that for a fair comparison, we use cuDNN to implement other layers (including 1$\times$1 convolution, pooling layers, etc.) and the $NCHW$ data layout (i.e., TVM's best-fit data layout) for both our solution and the TVM solution.

On A100, our five TKD-compressed models (DenseNet121, DenseNet201, ResNet18, ResNet50, and VGG16) using our optimized core convolution kernel with oracle/modeling achieves speedup of 2.14$\times$/2.11$\times$, 1.7$\times$/1.68$\times$, 
3.27$\times$/3.24$\times$, 2.07$\times$/ 2.04$\times$, and 2.37$\times$/2.33$\times$, respectively, compared with the original non-compressed models using cuDNN. The speedup over the TKD-compressed models using cuDNN is 1.41$\times$/1.38$\times$, 1.42$\times$/1.40$\times$, 2.21$\times$/2.18$\times$, 1.26$\times$/1.25$\times$, and 1.45$\times$/1.43$\times$, respectively.
The speedup over the TKD-compressed models using TVM is 1.03$\times$/1.01$\times$, 1.04$\times$/1.03$\times$, 1.12$\times$/1.11$\times$, 1.02$\times$/ 1.01$\times$, and 1.09$\times$/1.08$\times$, respectively.

On 2080Ti, the compressed models using our optimized core convolution kernel achieves speedup of 4.15$\times$/3.80$\times$, 2.62$\times$/2.55$\times$, 7.3$\times$/6.55$\times$, 2.83$\times$/2.75$\times$, and 2.73$\times$/2.45$\times$, respectively, compared with the non-compressed models implemented with cuDNN.
The speedup over the compressed models using cuDNN is 2.16$\times$/1.97$\times$, 1.81$\times$/1.74$\times$, 3.71$\times$/3.25$\times$, 1.38$\times$/1.35$\times$, and 1.68$\times$/1.53$\times$, respectively.
The speedup over the compressed models using TVM is 1.13$\times$/1.04$\times$, 1.13$\times$/ 1.08$\times$, 1.91$\times$/1.69$\times$, 1.09$\times$/1.06$\times$, 1.25$\times$/1.14$\times$, respectively.

\section{Conclusion and Future Work}
\label{sec:conclusion}

In this paper, we propose an efficient end-to-end framework to generate highly accurate and compact CNN models via Tucker decomposition and optimized C++/CUDA code for GPU inference. We incorporate hardware constraints into the model generation and develop a new scheme for Tucker-format convolutions with high-performance GPU kernels. 
Evaluation on modern CNNs with A100 demonstrates that our compressed models with optimized code achieve up to 2.37$\times$ speedup over cuDNN and 1.10$\times$ speedup over TVM. 

In the future, we plan to extend our work to cover wide CNNs such as GoogleNet and NasNet by developing a scheme that can determine the ranks for multiple concurrent convolutions and minimize the latency.

\section*{Acknowledgement}
\small 
The research was supported by the NSF awards 2232120, 2034169, 2042084, 1937403, 1955909, 2018016, and 2009007.
We thank Jacques Pienaar for helping us improve the paper quality.

\bibliographystyle{plain}
\bibliography{refs}

\clearpage
\newpage
\appendix

{%
\section{Artifact Description/Evaluation}
\label{artifacts-of-ppopp23-paper-tdc-towards-extremely-efficient-cnns-on-gpus-via-hardware-aware-tucker-decomposition}}

\hypertarget{download-artifacts}{%
\subsection{Overview}\label{download-artifacts}}
The artifacts contain all code, scripts, and datasets that are necessary for reproducing our results. The artifacts can be obtained from \url{https://github.com/black-cat-sheriff/TDC-PPOPP } and \url{https://doi.org/10.5281/zenodo.7439434}. 

\hypertarget{package-requirements}{%
\subsection{Package Requirements}\label{package-requirements}}

\begin{itemize}
\item
  cmake(\textgreater=3.10)
\item
  cuDNN(\textgreater=8)
\item
  CUDA \textgreater=11 (A100)
\item
  CUDA \textgreater=10 (2080Ti)
\item
  python3(\textgreater=3.7)
\item
  tensorly(0.7.4, pip install tensorly==0.7.0)
\item
  timm(0.5.4, pip install timm==0.5.4)
\item
  torch(\textgreater=1.4)
\item
  torchvision
\item
  numpy
\end{itemize}

Note that we require the version of timm to be 0.5.4.

\hypertarget{setup-very-important}{%
\subsection{SETUP}\label{setup-very-important}}

\begin{verbatim}
export CUDA_PREFIX='/usr/local/cuda'
export CUDA_INCLUDE=$CUDA_PREFIX/include
export CUDA_LIB64=$CUDA_PREFIX/lib64
export LD_LIBRARY_PATH=$CUDA_LIB64:$LD_LIBRARY_PATH
export PATH=$CUDA_PREFIX/bin:$PATH
\end{verbatim}

Please change \texttt{/usr/local/cuda} to the path to your CUDA library.
We assume that cuDNN is installed in \texttt{CUDA\_PREFIX}, which means
that cuDNN header files are in \texttt{CUDA\_INCLUDE} and cuDNN library
files are in \texttt{CUDA\_LIB64}. Also, please note that some cuDNN
versions only have \texttt{lib} folder rather than \texttt{lib64}
folder, please change \texttt{CUDA\_LIB64} accordingly.

\hypertarget{tdc-trained-model-evaluation}{%
\subsection{TDC TRAINED MODEL
EVALUATION}\label{tdc-trained-model-evaluation}}

\hypertarget{tucker-format-model-accuracy-evaluation.}{%
\subsubsection{Tucker-format model accuracy
evaluation Table \ref{tab:accuracy}.}\label{tucker-format-model-accuracy-evaluation.}}

We provided 6K images to demonstrate our models, but you are encouraged to obtain more images from ImageNet.
\begin{verbatim}
cd inference
python main.py --model tkc_resnet18  \
    --data-path test_images/
python main.py --model tkc_resnet50 \
    --data-path test_images/
python main.py --model tkc_densenet121 \
    --data-path test_images/
python main.py --model tkc_densenet201 \
    --data-path test_images/
python main.py --model tkc_vgg16 \
    --data-path test_images/
\end{verbatim}

Note that \texttt{test\_images} is the path to the folder where the
images used for the demo are saved (i.e., 6K images in total).

You will observe the following result.

\begin{figure}[H]
\centering
\includegraphics[width=\linewidth]{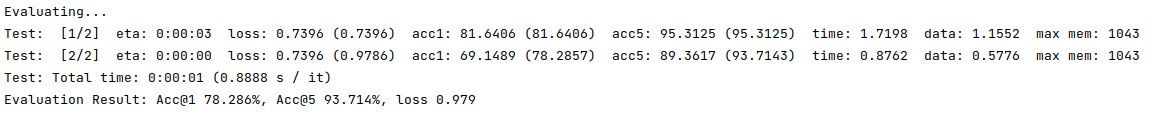}
\end{figure}

\hypertarget{performance-evaluation-of-tdc-generated-core-convolution-layers}{%
\subsubsection{Performance Evaluation of TDC-generated Core Convolution
Layers}\label{peformance-evaluation-of-tdc-generated-core-convolution-layers-on-2080ti}}

Please go to the main folder and run:

\begin{verbatim}
python3 run_2080Ti.py & python3 run_A100.py
\end{verbatim}

It will generate three tables.

\hypertarget{comparison-among-tdc-oracle-kernel-cudnn-and-tvm.}{%
\paragraph{A.4.2.1 Comparison among TDC oracle kernel, cuDNN, and
TVM.}\label{comparison-among-tdc-oracle-kernel-cudnn-and-tvm.}}

You will get the first table like below, including convolution shapes,
convolution schemes, runtimes, and TDC speedups. Please refer to Figure \ref{fig:2080Ti conv layer benchmark}.

\begin{figure}[H]
\centering
\includegraphics[width=\linewidth]{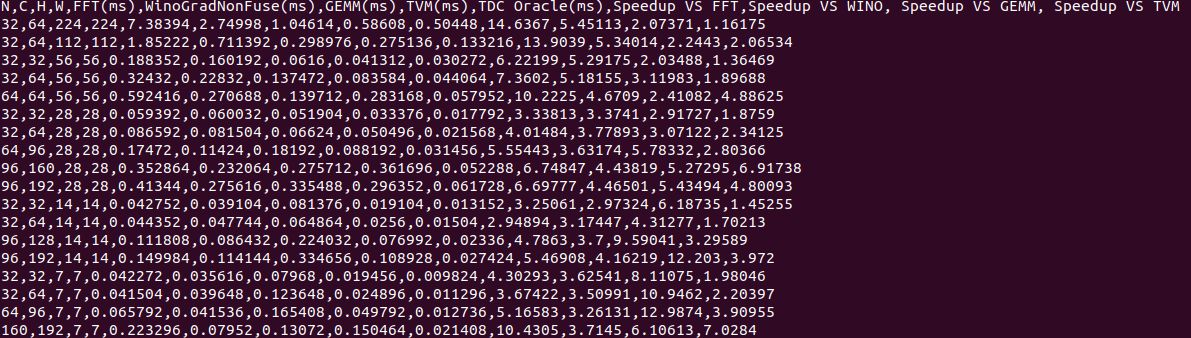}
\end{figure}

\hypertarget{comparison-among-tdc-modeling-kernel-cudnn-and-tvm.}{%
\paragraph{A.4.2.2 Comparison among TDC modeling kernel, cuDNN, and
TVM.}\label{comparison-among-tdc-modeling-kernel-cudnn-and-tvm.}}

You will get the second table like below, including convolution shapes,
convolution schemes, runtimes, and TDC speedups.

\begin{figure}[H]
\centering
\includegraphics[width=\linewidth]{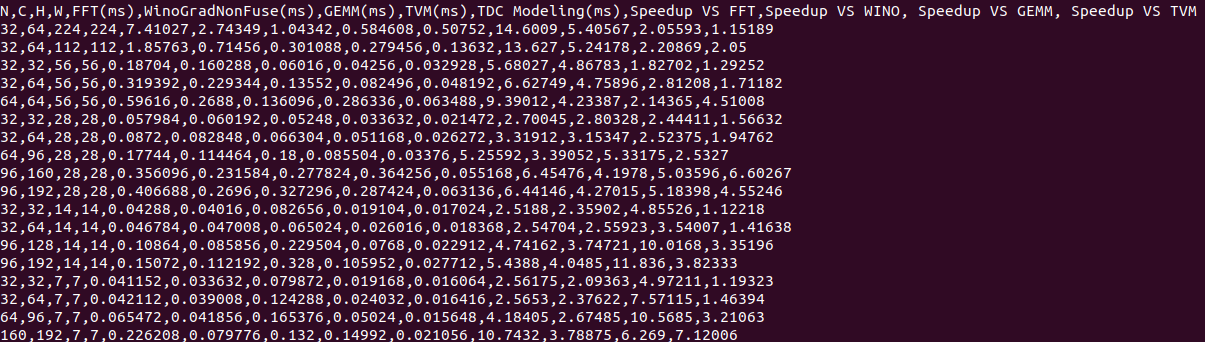}
\end{figure}

\hypertarget{end-to-end-performance-comparison-among-pure-cudnn-on-original-models-pure-cudnn-on-tk-compressed-models-and-tk-compressed-models.}{%
\paragraph{A.4.2.3 End-to-end performance comparison among pure cuDNN
on original models, pure cuDNN on TK compressed models, and TK
compressed
models.}\label{end-to-end-performance-comparison-among-pure-cudnn-on-original-models-pure-cudnn-on-tk-compressed-models-and-tk-compressed-models.}}

You will get the third table like below, each model has two rows - the
first one is header and the second one is runtime. Please refer to Figure \ref{fig:2080ti-end2end}.

\begin{figure}[H]
\centering
\includegraphics[width=\linewidth]{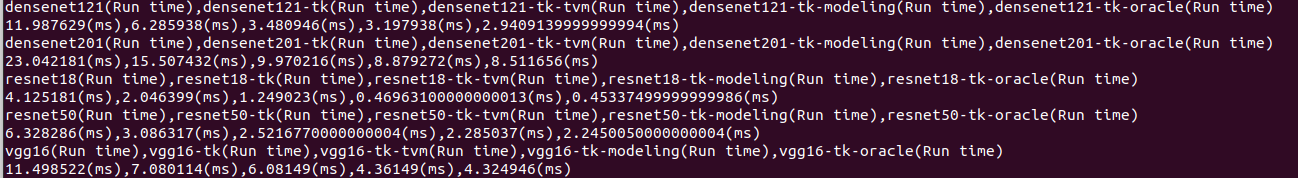}
\end{figure}



\end{document}